\documentclass[%
 reprint,
preprint,
bibnotes,
amsmath,amssymb,
onecolumn,
prc,
]{revtex4-1}

\usepackage{graphicx}
\usepackage{dcolumn}
\usepackage{bm}

\usepackage{enumitem}
\usepackage{amsmath}
\usepackage{amssymb}
\usepackage[nottoc,numbib]{tocbibind}
\settocbibname{References}

\begin{document}

\begin{figure} [htp]
\begin{center}
\includegraphics[height=10.0in]{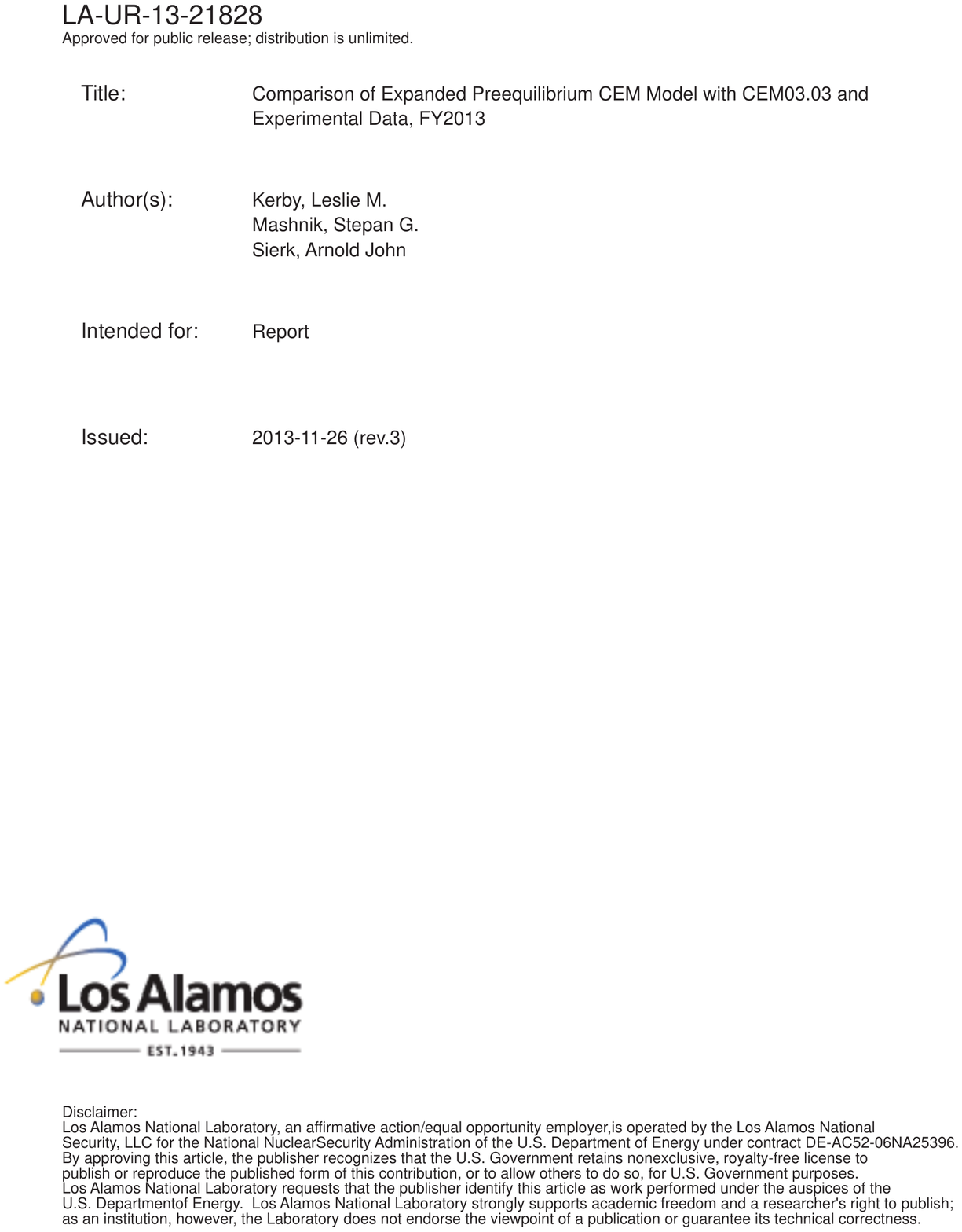}
\end{center}
\end{figure}

\clearpage

\title{Comparison of Expanded Preequilibrium CEM Model with CEM03.03 and Experimental Data}

\author{Leslie M. Kerby}
 \affiliation{University of Idaho, Idaho Falls, Idaho USA}
 \affiliation{Los Alamos National Laboratory, Los Alamos, New Mexico USA}

\author{Stepan G. Mashnik}
 \affiliation{Los Alamos National Laboratory, Los Alamos, New Mexico USA}

\author{Arnold J. Sierk}
 \affiliation{Los Alamos National Laboratory, Los Alamos, New Mexico USA}

\collaboration{LANL FY2013 Report, LA-UR-13-21828}

\date{September 25, 2013}

\maketitle

\newpage
\tableofcontents

\newpage

\section{Introduction}
Emission of light fragments (LF) from nuclear reactions is an open question. Different reaction mechanisms contribute to their production; the relative roles of each, and how they change with incident energy, mass number of the target, and the type and emission energy of the fragments is not completely understood.

None of the available models are able to accurately predict emission of LF from arbitrary reactions. However, the ability to describe production of LF (especially at energies $\gtrsim 30$~MeV) from many reactions is important for different applications, such as cosmic-ray-induced Single Event Upsets (SEUs), radiation protection, and cancer therapy with proton and heavy-ion beams, to name just a few. The Cascade-Exciton Model (CEM) \cite{CEMModel} version 03.03 and the Los Alamos version of the Quark-Gluon String Model (LAQGSM) \cite{LAQGSM, ICTP-IAEAWorkshop} version 03.03 event generators in Monte Carlo N-Particle Transport Code version 6 (MCNP6) \cite{MCNP6} describe quite well the spectra of fragments with sizes up to $^{4}$He across a broad range of target masses and incident energies (up to $\sim 5$~GeV for CEM and up to $\sim 1$~TeV/A for LAQGSM). However, they do not predict the high-energy tails of LF spectra heavier than $^4$He well. Most LF with energies above several tens of MeV are emitted during the precompound stage of a reaction. The current versions of the CEM and LAQGSM event generators do not account for precompound emission of LF larger than $^{4}$He. 

The aim of our work is to extend the precompound model in them to include such processes, leading to an increase of predictive power of LF-production in MCNP6. This entails upgrading the Modified Exciton Model currently used at the preequilibrium stage in CEM and LAQGSM. It will also include expansion and examination of the coalescence and Fermi break-up models used in the precompound stages of spallation reactions within CEM and LAQGSM. Extending our models to include emission of fragments heavier than $^4$He at the precompound stage has already provided preliminary results that have much better agreement with experimental data.

\section{Why This Research Is Needed}
\begin{figure} [htp]
\begin{center}
\includegraphics[height=2.0in]{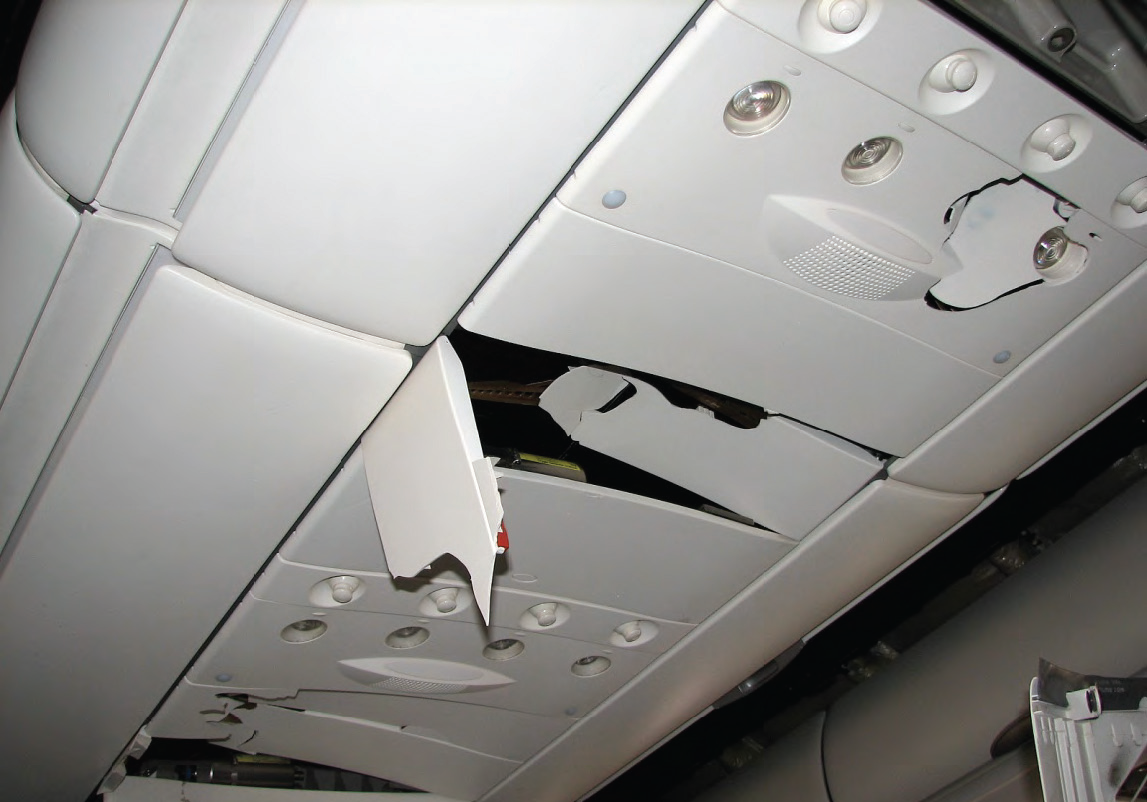}
\includegraphics[height=2.0in]{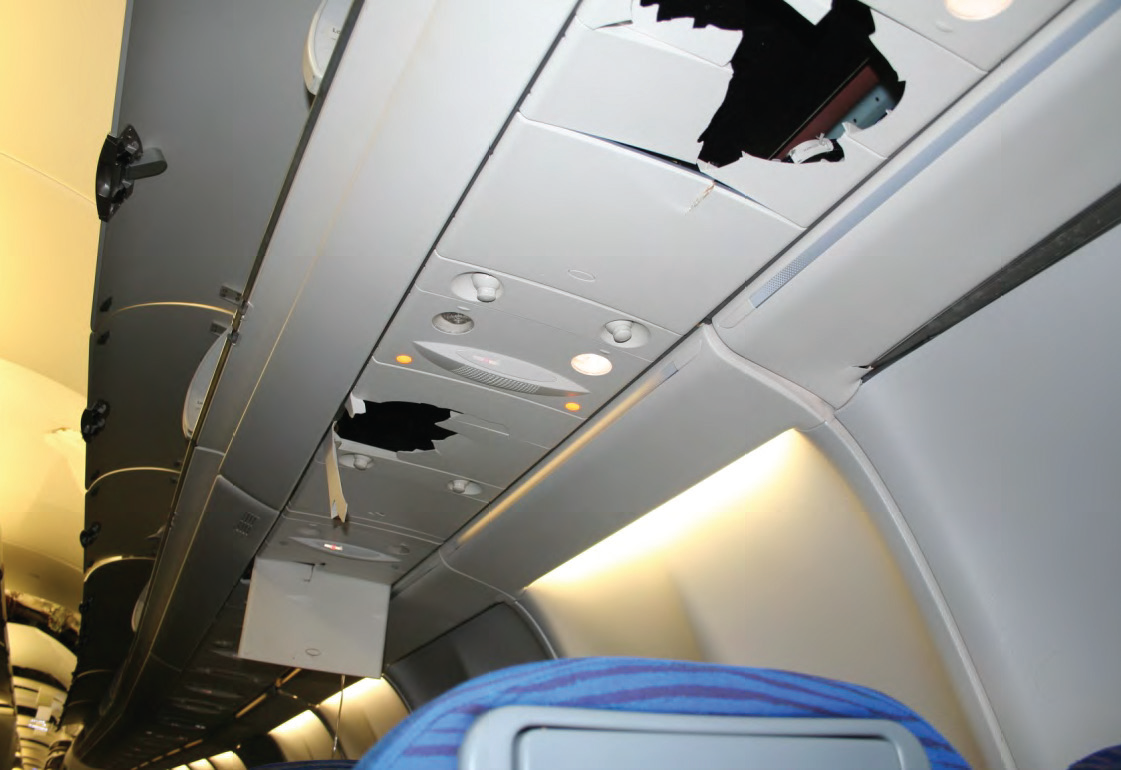}
\caption[]{Photographs of the damaged Airbus after the SEU \cite{NeciaGrantCooper}.}
\label{fig:Airbus}
\end{center}
\end{figure}

In October 2008 an Airbus plane was struck by a cosmic ray en route from Perth to Singapore, one of its inertial reference computer units failed, and it sharply lost altitude \cite{NeciaGrantCooper}. It did land safely, but as seen in Figure~\ref{fig:Airbus}, it caused significant injury to both the occupants and the plane. 

These SEUs are not rare, and can wreak significant havoc. For example, in a typical 14-day space mission the shuttles' 5 computers typically receive 400-500 SEUs \cite{Singleterry}. In addition, even though the plane accident was serious, much more serious incidents have occurred: during the Cold War a U. S. satellite was hit by a cosmic ray and reported that there had been a nuclear missile launch, heading toward the U. S. \cite{CountdownToZero}. The U. S. went on high alert and readied their nuclear weapons. Thankfully they were never launched. Understanding how high-energy fragments interact with matter is critical to preventing these malfunctions.

Accurate simulation of LF spectra is also important in the fields of radiation shielding, especially for applications in space. Modern computers cannot be used in space because the electronics are too small and delicate and cannot, at present, be shielded well enough. An even larger problem is radiation shielding for the human astronauts exposed to Galactic Cosmic Rays (GCRs) \cite{Singleterry}.

This research is also important to several medical fields, such as cancer treatment with proton or heavy-ion beams. Proton and heavy-ion therapy has been shown to be more effective than x-ray therapy, and have much fewer side effects \cite{Protons}.

Another indication of the importance of this research is the recommendation of an international evaluation and comparison, the 2008-2010 IAEA (International Atomic Energy Agency) Benchmark of Spallation Models, that we make this change in our code \cite{SecondAdvancedWorkshop,IAEABenchmark}. While no other spallation model can generally predict high-energy light fragment emission from arbitrary reactions, it is an accomplishment several model development groups are working to achieve.

Furthermore, MCNP6's GENXS option at present does not produce tallies for particles larger than $^{4}$He. This limitation is serious for some of our interest groups. For example, NASA recently contacted one of us (SGM) to inquire if our codes could produce LF spectra in the intermediate- and high-energy regimes. At present they cannot.

Last, but not least, this research helps us understand better the mechanisms of nuclear reactions. 

\section{Current Capabilities of CEM03.03}
\subsection{Overview of the CEM Model}
\begin{figure}[htp]
\centering
\includegraphics[width=4.5in]{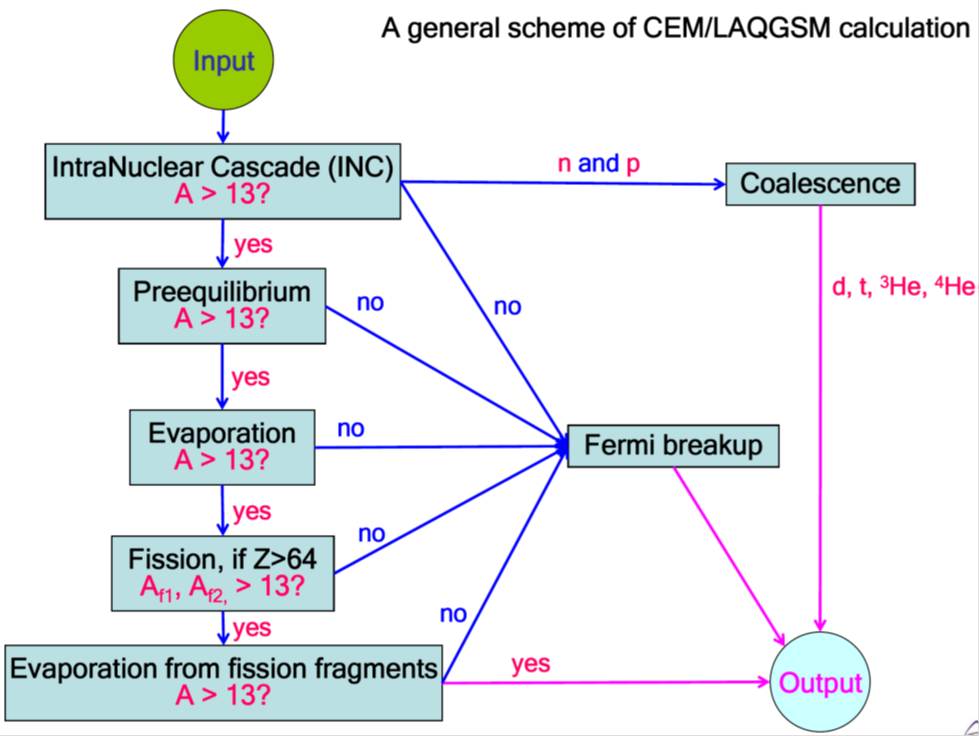}
\caption[]{Flowchart of nuclear-reaction calculations by CEM03.03 \cite{ICTP-IAEAWorkshop}.}
\label{fig:Flowchart}
\end{figure}

As a rule, a reaction begins with the IntraNuclear Cascade, referred to as either the INC or as the Cascade (see Fig.~\ref{fig:Flowchart}). The incident particle or nucleus (in the case of using LAQGSM) enters the target nucleus and begins interacting with nucleons, scattering off them and also often creating new particles in the process. The incident particle and all newly created particles are followed until they either escape from the nucleus or reach a threshold energy (roughly 10-30 MeV per nucleon) and are then considered ``absorbed" by the nucleus. 

The preequilibrium stage uses the Modified Exciton Model (MEM) to determine emission of protons, neutrons, and fragments up to $^4$He from the residual nucleus. We discuss the MEM in more detail below. This stage can have a highly excited residual nucleus undergoing dozens of exciton transitions and particle emissions. The preequilibrium stage ends when the residual nucleus is just as likely to have a $\Delta n = +2$ exciton transaction as a $\Delta n = -2$ exciton transaction.

In the evaporation stage neutrons and protons in the outer shells of the residual nucleus can ``evaporate" off, either singly or as fragments. The CEM evaporation stage is modeled after Furihata's Generalized Evaporation Model (GEM2) \cite{GEM2}, and can emit light fragments up to $^{28}$Mg.

During and after evaporation, the code looks to see if we have an isotope that has $Z \geq 65$ and is fissionable. If it is, and there is fission, then the code follows the evaporation stage for the fission fragments.

There are two models that are not directly part of this linear progression: Coalescence and Fermi break-up (see Fig.~\ref{fig:Flowchart}). The Cascade stage only emits neutrons, protons, and pions (and other particles, in the case of using LAQGSM at high energies), so the coalescence model ``coalesces" some of the neutrons and protons produced during the INC into larger fragments, by comparing their momenta. If their momenta are similar enough then they coalesce. The current coalescence model can only coalesce up to a $^4$He fragment, the same as the preequilibrium stage. The Fermi break-up is an oversimplified multifragmentation model that is fast and accurate for small atomic numbers, so we use it when the residual mass number is less than or equal to 13.

\subsection{Comparison with Experimental Data by Machner et al.}
\begin{figure}[htp]
\centering
\includegraphics[trim = 0.5in 3.5in 1.0in 3.5in, width=6.0in]{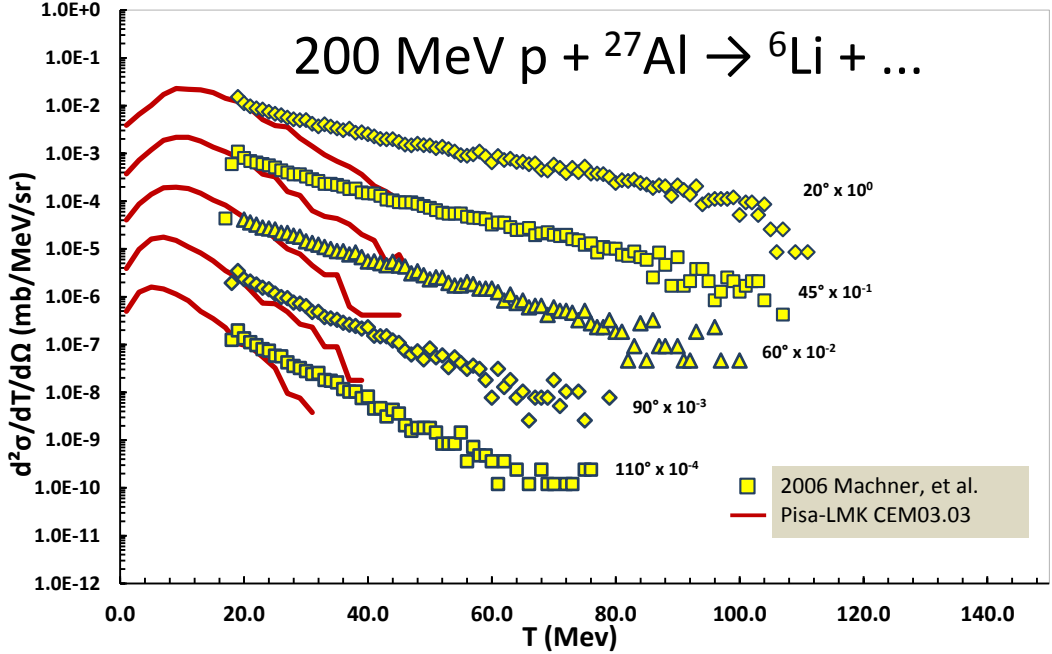}
\caption[]{Comparison of CEM03.03 (solid red lines) and experimental data by Machner et al. \cite{Machner} (open points).}
\label{fig:p200AlCompOld}
\end{figure}

Figure~\ref{fig:p200AlCompOld} shows the double-differential cross section of the reaction 200 MeV p + $^{27}$Al $\rightarrow$ $^{6}$Li, comparing Machner et al. \cite{Machner} experimental data (open points) and unmodified CEM03.03 (solid red lines). 

The vertical axis presents the double differential cross sections. The horizontal axis shows the kinetic energy of the emitted particles ($^6$Li in this case) in MeV. The different data bands represent $^6$Li detected (or simulated) at different angles, and are separated out by multiplying each band by a different factor of 10. As can be seen, the current version of CEM does not predict the high-energy tails of $^6$Li well. This is true across other reaction energies and target mass numbers for all fragments heavier than $^4$He, for higher energies. At lower energies ($\lesssim 25$~MeV) CEM matches well, but as we enter intermediate energies ($\gtrsim 25$~MeV) CEM falls off sharply. This is because the peak which occurs at lower energy is a result of the evaporative stage, which does consider emission of LF (up to $^{28}$Mg). The intermediate section of the fragment spectra tail (up to $\sim 150$~MeV) is largely produced by the MEM within the preequilibrium stage. The higher energy tail of fragment spectra is largely produced from coalescence, also a precompound stage. Neither the MEM nor the coalescence model presently consider emission of light fragments heavier than $^4$He.

\section{Emission of High-Energy LF in Other Models}
This paper focuses on the emission of high-energy LF at the preequilibrium stage of nuclear reactions. However, high-energy LF can be produced at other stages of reactions. So, Cugnon {\it et al.} have modified their Li\`{e}ge IntraNuclear Cascade (INCL) code to consider emission of light fragments heavier than $^4$He during the cascade stage of reactions via coalescence of several nucleons at the nuclear periphery \cite{Cugnon}. These modifications have not yet been generalized across all types of reactions. In addition, the INCL+ABLA model is limited to relatively light incident projectiles (particles and light ions, typically, up to oxygen). Several previous papers by the same group discuss the production of light fragments up to $A=10$ (see, e.g., \cite{Cugnon2010, Cugnon2011}). A recent 2013 paper by the same authors presents satisfactory results for emission spectra of $^6$He, $^6$Li, $^7$Li, and $^7$Be in the reaction $p + ^{197}Au \rightarrow ...$ and discusses emission of clusters up to $A = 12$ \cite{INCL4.6}. 

Emission of $^7$Be at the preequilibrium stage (described by a hybrid exciton model and coalescence pick-up model) was studied by A. Yu. Konobeyev and Yu. A. Korovin more than a decade ago \cite{Konobeyev}. Additionally, preequilibrium emission of helium and lithium ions and the necessary adjustments to the Kalbach systematics was discussed in Ref. \cite{Uozumi}. Preequilibrium emission of light fragments was also studied within the CEM in 2002 \cite{CEM2k2f}, but that project was never completed. 

Finally, energetic fragments can be produced via Fermi break-up \cite{Fermi} and multifragmentation processes, as described, e.g., by the Statistical Multifragmentation Model (SMM) \cite{SMM}; (see a comparison of the Fermi break-up model with SMM in the recent paper by Souza {\it et al.} \cite{Souza2013}).

Light fragments can also be emitted during the compound stage of  reactions. GEM2, the evaporation model used in CEM, emits light fragments up to $^{28}$Mg \cite{GEM2}. In addition, light fragments can be produced via very asymmetric binary fission, as described, e.g., by the fission-like binary decay code GEMINI by Charity et al. \cite{Charity01}, and also via ternary fission. For more information, see the recent Ref. \cite{Ronen} wherein Y. Ronen discusses the physics of how light fragments are products seen in ternary fission. However, neither evaporation nor fission processes can produce high-energy fragments, of interest to our current study.

\section{The Modified Exciton Model (MEM)}
\subsection{MEM Code}
\begin{figure}[htp]
\centering
\includegraphics[width=5.5in]{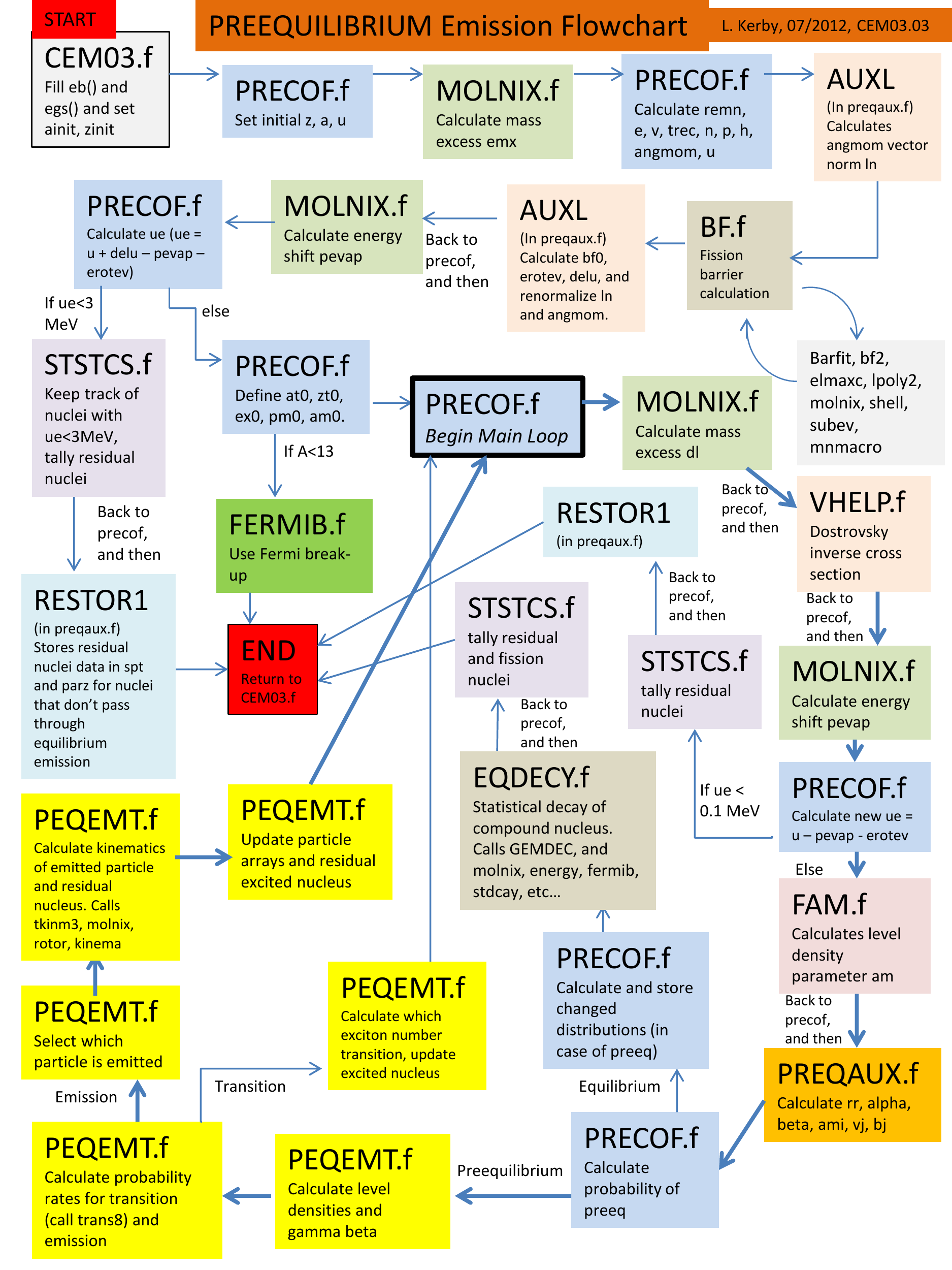}
\caption[]{Flowchart for emission of light fragments in the MEM code.}
\label{fig:MEMFlowchart}
\end{figure}

Let us present below an in-depth description of the code in MEM calculations. The flowchart in Figure~\ref{fig:MEMFlowchart} describes the calculations and processes performed in the MEM. 

\subsection{MEM Physics}
The probability of finding the system at the time moment $t$ in the $E\alpha$ state, $P(E,\alpha,t)$, is given by the following differential equation:
\begin{equation}
\frac{\delta P(E,\alpha,t)}{\delta t} = \sum_{\alpha \neq \alpha '}[\lambda(E\alpha,E\alpha')P(E,\alpha',t) - \lambda(E\alpha',E\alpha)P(E,\alpha,t)] .
\label{Master}
\end{equation}
Here $\lambda(E\alpha,E\alpha')$ is the energy-conserving probability rate, defined in the first order of the time-dependent perturbation theory as
\begin{equation}
\lambda (E\alpha,E\alpha') = \frac{2\pi}{h} |<E\alpha|V|E\alpha'>|^2 \omega_{\alpha}(E) .
\label{LambdaGeneral}
\end{equation}
The matrix element $<E\alpha|V|E\alpha'>$ is believed to be a smooth function in energy, and $\omega_\alpha(E)$ is the density of the final state of the system. One should note that Eq.~(\ref{Master}) is derived provided that the ``memory" time $\tau_{mem}$ of the system is small compared to the characteristic time for intranuclear transition $~\frac{\hbar}{\lambda(E\alpha,E\alpha')}$ but, on the other hand, Eq.~(\ref{Master}) itself is applicable for the time moments $t \gg \frac{\hbar}{\lambda(E\alpha,E\alpha')}$. Due to the condition $\tau_{mem} \gg \frac{\hbar}{\lambda(E\alpha,E\alpha')}$, being described by Eq.~(\ref{Master}), the random process is the Markovian one.

The Modified Exciton Model (MEM) \cite{CEMModel, Gudima, MODEX} utilized by CEM and LAQGSM uses effectively the relationship of the master equation (\ref{Master}) with a Markovian random processes. Indeed, an attainment of the statistical equilibration described by Eq.~(\ref{Master}) is an example of the discontinuous Markovian process: the temporal variable changes continuously and at a random moment the state of the system changes by a discontinuous jump, the behavior of the system at the next moment being completely defined by its state at present. As long as the transition probabilities $\lambda(E\alpha,E\alpha')$ are time independent, the waiting time for the system in the $E\alpha$ state has the exponential distribution (the Poisson flow) with the average lifetime $\frac{\hbar}{\Lambda(\alpha,E)} = \frac{\hbar}{\sum_\alpha'{\lambda(E\alpha,E\alpha')}}$. This fact prompts a simple method of solving the related system of Eq.~(\ref{Master}): simulation of the random process by the Monte Carlo technique. In this treatment it is possible to generalize the exciton model to all nuclear transitions with $\Delta n = 0, \pm 2$, and the multiple emission of particles and to depletion of nuclear states due to the particle emission. In this case the system (\ref{Master}) is as follows:~\cite{Mashnik1994}
\begin{equation}
\begin{split}
\frac{\delta P(E,\alpha,t)}{\delta t} = & -\Lambda(n,E)P(E,n,t) + \lambda_+(n-2,E)P(E,n-2,t) + \\
	& + \lambda_0(n,E)P(E,n,t) + \lambda_-(n+2,E)P(E,n+2,t) + \\
	& + \sum_j \int dT \int dE' \lambda_c^j (n,E,T)P(E',n+n_j,t)\delta(E' -E-B_j-T) .
\end{split}
\label{Probability}
\end{equation}
Now we solve our master equation Eq.~(\ref{Probability}) by finding the particle emission rates $\lambda_c^j$ and the exciton transition rates $\lambda_+$, $\lambda_0$, and $\lambda_-$.

\vspace*{0.25cm}
{\noindent \bf \em Particle Emission} \\
According to the detailed balance principle, the emission  width $\Gamma _{j}$, (or probability of emitting particle fragment $j$), is estimated as
\begin{equation}
\Gamma_{j}(p,h,E) = \int_{V_j^c}^{E-B_j} \lambda_c^j (p,h,E,T)dT ,
\end{equation}
where the partial transmission probabilities, $\lambda_c^j$, are equal to
\begin{equation}
\lambda_c^j (p,h,E,T) = \frac{2s_j + 1}{\pi^2\hbar^3} \mu_j \Re (p,h) \frac{\omega (p-1,h,E-B_j-T)}{\omega (p,h,E)} T \sigma_{inv} (T) .
\label{LambdaTransmission}
\end{equation}

\begin{itemize}[noitemsep]
	\item[] $s_j$: spin of the emitted particle $j$ 
	\item[] $\mu_j$: reduced mass of the emitted particle $j$ 
	\item[] $\omega$: level density of the $n$-exciton state 
	\item[] $B_j$: binding energy 
	\item[] $V_j^c$: Coulomb barrier 
	\item[] $T$: kinetic energy of the emitted particle $j$
	\item[] $\sigma_{inv}$: inverse cross section
	\item[] $\Re$: creates zero probability of emission if the number of particle excitons is less than the number nucleons of particle $j$ 
\end{itemize}

Equation~(\ref{LambdaTransmission}) describes the emission of neutrons and protons. For complex particles, the level density formula $\omega$ becomes more complicated and an extra factor $\gamma_j$ must be introduced:
\begin{equation}
\gamma_j \approx p_j^3 (\frac{p_j}{A})^{p_j - 1} .
\label{GammaBeta}
\end{equation}

In reality Equation~(\ref{GammaBeta}) for $\gamma_j$ is a preliminary rough estimation that is refined by parameterizing over a mesh of residual nuclei energy and mass number \cite{CEMUserManual}. Adding the possibility of LF emission alters the previous parameterization, effectively requiring new parameterization. This work of parameterizing $\gamma_j$ still needs to be done in order to generalize our results to all energies and target masses. In addition, we would like to add better modeling of $\gamma_j$: investigating the use of physical models and/or adding extrapolation to the mesh. 

Assuming an equidistant level scheme with the single-particle density $g$, we have the level density of the $n$-exciton state as~\cite{Ericson}
\begin{equation}
\omega(p,h,E) = \frac{g (gE)^{p+h-1}}{p! h! (p+h-1)!} \mbox{ .}
\label{OmegaGeneral}
\end{equation}
This expression should be substituted into Eq.~\ref{LambdaTransmission} to obtain the transmission rates $\lambda_c^j$. 

\vspace*{0.5cm}
{\noindent \bf \em Exciton Transitions} \\
According to Equation~(\ref{LambdaGeneral}), for a preequilibrium nucleus with excitation energy $E$ and number of excitons $n=p+h$, the partial transition probabilities changing the exciton number by $\Delta n$ are
\begin{equation}
\lambda_{\Delta n} (p,h,E) =
\frac{2\pi}{\hbar}|M_{\Delta n}|^2 \omega_{\Delta n} (p,h,E) \mbox{ .}
\label{LambdaTransitionGeneral}
\end{equation}
For these transition rates, one needs the number of states, $\omega$, taking into account the selection rules for intranuclear exciton-exciton scattering. The appropriate formulae have been derived by Williams~\cite{Williams} and later corrected for the exclusion principle and indistinguishability of identical excitons in Refs.~\cite{Williams2,Ribansky}:
\begin{eqnarray}
\omega _+ (p,h,E) & = &
\frac{1}{2} g \frac{[gE-{\cal A}(p+1,h+1)]^2}
{n+1}
\biggl[ \frac{gE - {\cal A}(p+1,h+1)}{gE - {\cal A}(p,h)} \biggr] ^{n-1}
\mbox{ ,} \nonumber \\
\omega _0 (p,h,E) & = &
\frac{1}{2} g \frac{[gE-{\cal A}(p,h)]}{n} [p(p-1)+4ph+h(h-1)]
\mbox{ ,} \nonumber \\
\omega _- (p,h,E) & = &
\frac{1}{2} gph(n-2) \mbox{ ,}
\label{OmegaTransition}
\end{eqnarray}
where ${\cal A}(p,h) = (p^2 +h^2 +p-h)/4 - h/2$. By neglecting the difference of matrix elements with different $\Delta n$, $M_+ = M_- = M_0 = M$, we estimate the value of $M$ for a given nuclear state by associating the $\lambda_+ (p,h,E)$ transition with the probability for quasi-free scattering of a nucleon above the Fermi level on a nucleon of the target nucleus. Therefore, we have
\begin{equation}
\frac{ < \sigma (v_{rel}) v_{rel} >}{V_{int}} =
\frac{\pi}{\hbar} |M|^2 
\frac{g [ gE-{\cal A}(p+1,h+1)]}{n+1}
\biggl[ \frac{gE - {\cal A}(p+1,h+1)}{gE - {\cal A}(p,h)} \biggr] ^{n-1}
\mbox{ .}
\label{SigmaAverage}
\end{equation}
Here, $V_{int}$ is the interaction volume estimated as $V_{int} = {4 \over 3} \pi (2 r_c + \lambda / 2 \pi)^3$, with the de Broglie wave length $\lambda / 2 \pi$ corresponding to the relative velocity $v_{rel} = \sqrt{2 T_{rel} /m_N}$. A value of the order of the nucleon radius is used for $r_c$ in the CEM: $r_c = 0.6$ fm.

The averaging on the left-hand side of Eq.~(\ref{SigmaAverage}) is carried out over all excited states, taking into account the exclusion principle. Combining~(\ref{LambdaTransitionGeneral}), (\ref{OmegaTransition}), and (\ref{SigmaAverage}) we finally get for the transition rates:
\begin{eqnarray}
\lambda _+ (p,h,E) & = &
\frac{ < \sigma (v_{rel}) v_{rel} >}{V_{int}} \mbox{ ,} \nonumber \\
\lambda _0 (p,h,E) & = &
\frac{ < \sigma (v_{rel}) v_{rel} >}{V_{int}}
\frac{n+1}{n}
\biggl[ \frac{gE - {\cal A}(p,h)}{gE - {\cal A}(p+1,h+1)} \biggr] ^{n+1}
\frac{p(p-1)+4ph+h(h-1)}{gE-{\cal A}(p,h)} \mbox{ ,} \nonumber \\
\lambda _- (p,h,E) & = &
\frac{ < \sigma (v_{rel}) v_{rel} >}{V_{int}}
\biggl[ \frac{gE - {\cal A}(p,h)}{gE - {\cal A}(p+1,h+1)} \biggr] ^{n+1}
\frac{ph(n+1)(n-2)}{[gE-{\cal A}(p,h)]^2} \mbox{ .}
\label{LambdaTransition}
\end{eqnarray}

\vspace*{0.25cm}
{\noindent \bf \em Angular Distributions} \\
The CEM predicts forward peaked (in the laboratory system) angular distributions for preequilibrium particles. For instance, CEM03.03 assumes that a nuclear state with a given excitation energy $E^*$ should be specified not only by the exciton number $n$ but also by the momentum direction $\Omega$. Following Ref.~\cite{Mantzouranis}, the master equation (Eq.~(\ref{Probability})) can be generalized for this case provided that the angular dependence for the transition rates $\lambda _+$, $\lambda _0$, and $\lambda _-$ (Eq.~(\ref{LambdaTransition})) is factorized. In accordance with Eq.~\ref{SigmaAverage}, in the CEM it is assumed that
\begin{equation}
<\sigma> \to <\sigma> F(\Omega) \mbox{ ,}
\label{SigmaFactor}
\end{equation}
where
\begin{equation}
F(\Omega) = {d \sigma^{free}/ d \Omega \over
\int d \Omega '  d \sigma^{free} / d \Omega '} \mbox{ .} 
\label{Factor}
\end{equation}
The scattering cross section $ d \sigma^{free}/ d \Omega$ is assumed to be isotropic in the reference frame of the interacting excitons, thus resulting in an asymmetry in both the nucleus center-of-mass and laboratory frames. The angular distributions of preequilibrium complex particles are assumed to be similar to those for the nucleons in each nuclear state \cite{CEMModel}.

This calculational scheme is easily realized by the Monte-Carlo technique. It provides a good description of double-differential spectra of preequilibrium nucleons and a not-so-good but still satisfactory description of complex-particle spectra from different types of nuclear reactions at incident energies from tens of MeV to several GeV. For incident energies below about 200 MeV, Kalbach \cite{Kalbach88} has developed a phenomenological systematics for preequilibrium-particle angular distributions by fitting available measured spectra of nucleons and complex particles. As the Kalbach systematics are based on measured spectra, they describe very well the double-differential spectra of preequilibrium particles and generally provide a better agreement of calculated preequilibrium complex particle spectra with data than does the CEM approach based on Eqs.~(\ref{SigmaFactor},\ref{Factor}). This is why we have incorporated into CEM03.03 the Kalbach systematics \cite{Kalbach88} to describe angular distributions of both preequilibrium nucleons and complex particles at incident energies up to 210 MeV. At higher energies, we use in CEM03.03 the CEM approach based on Eqs.~(\ref{SigmaFactor},\ref{Factor}).

\subsection{Precompound Particles Considered}

\begin{table}[here]
\caption{The emitted particles considered by the modified MEM}
\begin{ruledtabular}
\begin{tabular}{rlllllll}
\hline\hline 
 $Z_j$\hspace{2mm} & \multicolumn{7}{l} {Ejectiles} \\
\hline
0\hspace{2mm}  & n       &         &         &         &         &         &         \\
1\hspace{2mm}  & p       &\hspace{1mm}   d     &\hspace{1mm}   t     &         &         &         &         \\
2\hspace{2mm}  &$^{3 }$He&\hspace{1mm}$^{4 }$He&\hspace{1mm}$^{6 }$He&\hspace{1mm}$^{8 }$He&         &         &         \\
3\hspace{2mm}  &$^{6 }$Li&\hspace{1mm}$^{7 }$Li&\hspace{1mm}$^{8 }$Li&\hspace{1mm}$^{9 }$Li&         &         &         \\
4\hspace{2mm}  &$^{7 }$Be&\hspace{1mm}$^{9 }$Be&\hspace{1mm}$^{10}$Be&\hspace{1mm}$^{11}$Be&\hspace{1mm}$^{12}$Be&         &         \\
5\hspace{2mm}  &$^{8 }$B &\hspace{1mm}$^{10}$B &\hspace{1mm}$^{11}$B &\hspace{1mm}$^{12}$B &$\hspace{1mm}^{13}$B &         &         \\
6\hspace{2mm}  &$^{10}$C &\hspace{1mm}$^{11}$C &\hspace{1mm}$^{12}$C &\hspace{1mm}$^{13}$C &\hspace{1mm}$^{14}$C &\hspace{1mm}$^{15}$C &\hspace{1mm}$^{16}$C \\
7\hspace{2mm}  &$^{12}$N &\hspace{1mm}$^{13}$N &\hspace{1mm}$^{14}$N &\hspace{1mm}$^{15}$N &\hspace{1mm}$^{16}$N &\hspace{1mm}$^{17}$N &         \\
8\hspace{2mm}  &$^{14}$O &\hspace{1mm}$^{15}$O &\hspace{1mm}$^{16}$O &\hspace{1mm}$^{17}$O &\hspace{1mm}$^{18}$O &\hspace{1mm}$^{19}$O &\hspace{1mm}$^{20}$O \\
9\hspace{2mm}  &$^{17}$F &\hspace{1mm}$^{18}$F &\hspace{1mm}$^{19}$F &\hspace{1mm}$^{20}$F &\hspace{1mm}$^{21}$F &         &         \\
10\hspace{2mm} &$^{18}$Ne&\hspace{1mm}$^{19}$Ne&\hspace{1mm}$^{20}$Ne&\hspace{1mm}$^{21}$Ne&\hspace{1mm}$^{22}$Ne&\hspace{1mm}$^{23}$Ne&\hspace{1mm}$^{24}$Ne\\
11\hspace{2mm} &$^{21}$Na&\hspace{1mm}$^{22}$Na&\hspace{1mm}$^{23}$Na&\hspace{1mm}$^{24}$Na&\hspace{1mm}$^{25}$Na&         &         \\
12\hspace{2mm} &$^{22}$Mg&\hspace{1mm}$^{23}$Mg&\hspace{1mm}$^{24}$Mg&
\hspace{1mm}$^{25}$Mg&\hspace{1mm}$^{26}$Mg&\hspace{1mm}$^{27}$Mg&
\hspace{1mm}$^{28}$Mg\\
\hline\hline 
\end{tabular}
\end{ruledtabular}
\label{Particles}
\end{table}

Table~\ref{Particles} displays the particles our expanded MEM is designed to emit. Our model has been expanded to emit all 66 of these isotopes (through $^{28}$Mg).

\section{Results}

\subsection{Code Crash Protection}
Bugs used to be fixed on an as-encountered basis. However, after encountering one bug that could not feasibly be fixed in this manner, we decided to complete CEM-wide code crash protection. The entirety of the CEM code was modified to check, by if statements, for divide-by-zero errors and, if encountered, output error statements revealing where in the code such errors occurred (while fixing the divide-by-zero error to allow for completion of the simulations). Square root calculations were also protected to ensure no errors occurred. Logarithmic and inverse trigonometric functions were not error protected. 

\begin{figure}[]
\centering
\includegraphics[trim = 0.4in 1.5in 0.5in 1.5in, width=6.5in]{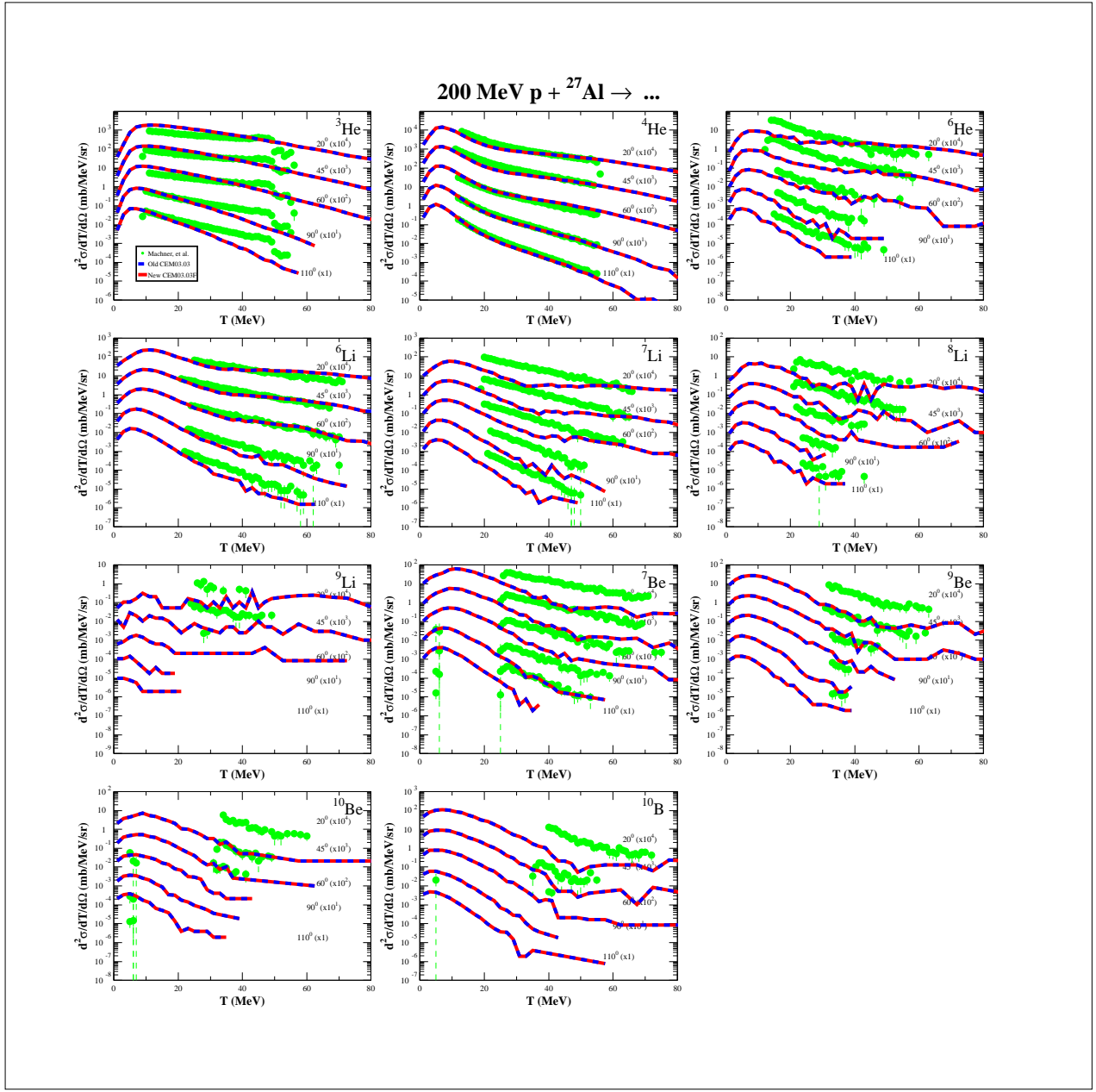}
\caption[]{Comparison of experimental data by Machner {\it et al.} \cite{Machner} (green points) with results from the Before-CCP CEM03.03 (blue dotted lines) and the After-CCP CEM03.03 (red solid lines) for 200 MeV p + $^{27}$Al $\rightarrow$ $...$ The before and after code crash protection results are equivalent.}
\label{fig:p200AlCCPComp}
\end{figure}

This was a large project as it involved slight modification of all the CEM code. However, as it will provide crash protection for future applications of CEM, including crash protection within future versions of MCNP, we determined it was worth it. 

As this crash protection involved the addition of numerous if-statements into the code, we investigated the impact on computation time. The influence on CPU runtime was not significant and could not be detected above the normal variations in runtime that occur due to time-of-day CPU speed fluctuations, or having a month between runs (and LANL servers subsequently getting faster, perhaps). In addition, we validated the crash protected code by rerunning many reactions to ensure we got the same results as the non-protected code. Fig.~\ref{fig:p200AlCCPComp} is an example of the before and after results. As can be seen, they are identical and the ``before'' blue dashed line is not even visible underneath the ``after'' solid red line.

\subsection{Recalibration of MEM Parameters}
With the expansion of the preequilibrium model to allow emission of light fragments up to $^{28}$Mg complete, we then turned our attention to recalibrating $\gamma_{\beta}$. This process is long and involves the re-fitting of all available reliable experimental data. We are in the middle of this process, but include several results below. Preliminary results are very encouraging.

\subsubsection{200 MeV p + $^{27}$Al}
\begin{figure}[]
\centering
\includegraphics[trim = 0.4in 1.5in 0.5in 1.5in, width=6.5in]{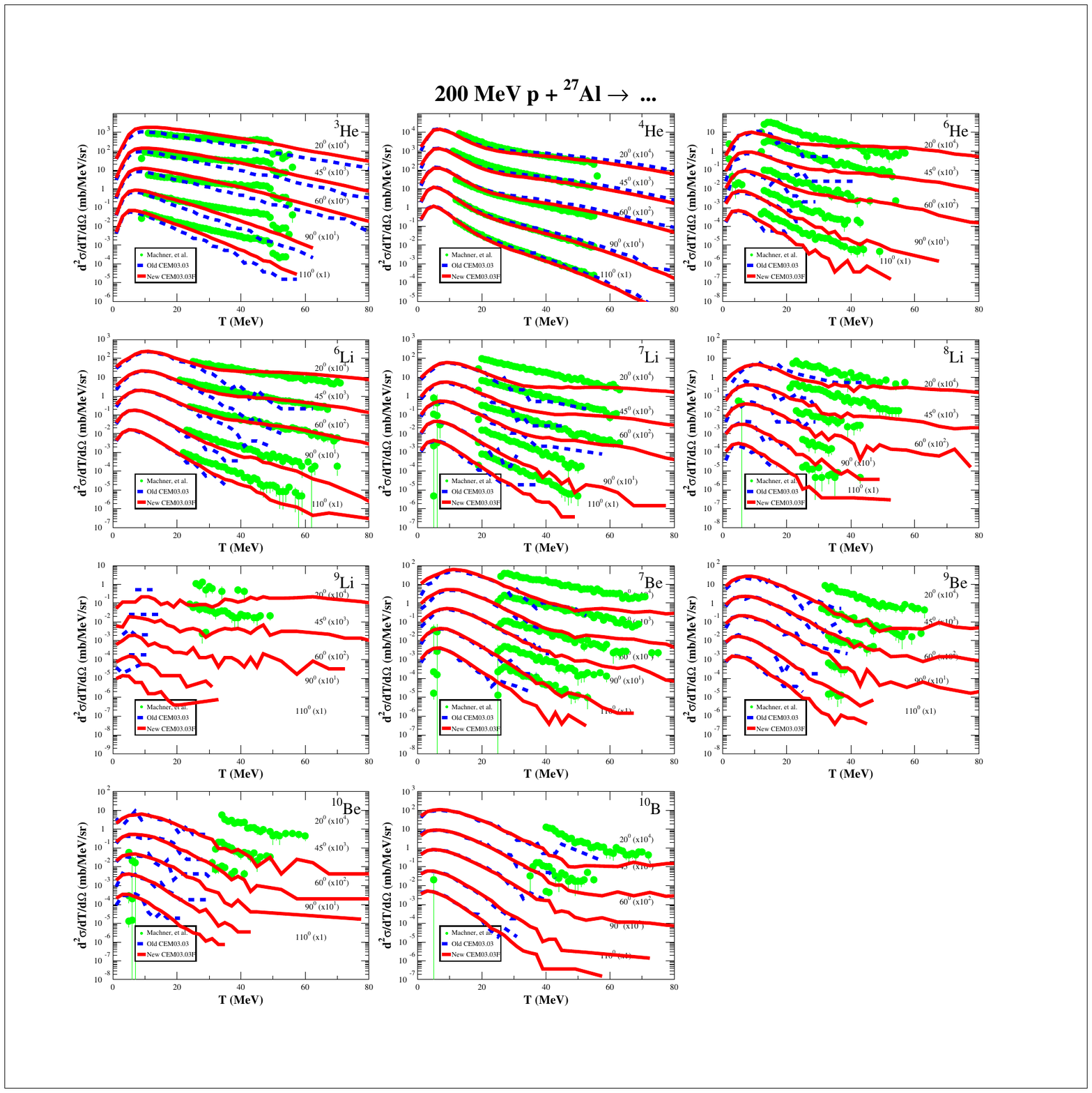}
\caption[]{Comparison of experimental data by Machner {\it et al.} \cite{Machner} (green points) with results from the unmodified CEM03.03 (blue dotted lines) and the modified MEM CEM03.03 (red solid lines) for 200 MeV p + $^{27}$Al $\rightarrow$ $...$}
\label{fig:p200Al}
\end{figure}

Figure~\ref{fig:p200Al} demonstrates the potential of the modified precompound code we built, for the same reaction and data as shown in Figure~\ref{fig:p200AlCompOld}, 200 MeV p + $^{27}$Al. The red solid lines show results from the new precompound code we designed in FY2013; the blue dotted lines present calculations from the old code; and the green points are experimental data from Machner, {\it et al.} \cite{Machner}. The upgraded MEM provides dramatically improved ability to describe the cross section at intermediate to high energies. 

\begin{figure}[]
\centering
\includegraphics[trim = 0.5in 0.5in 1.0in 1.0in, width=6.5in]{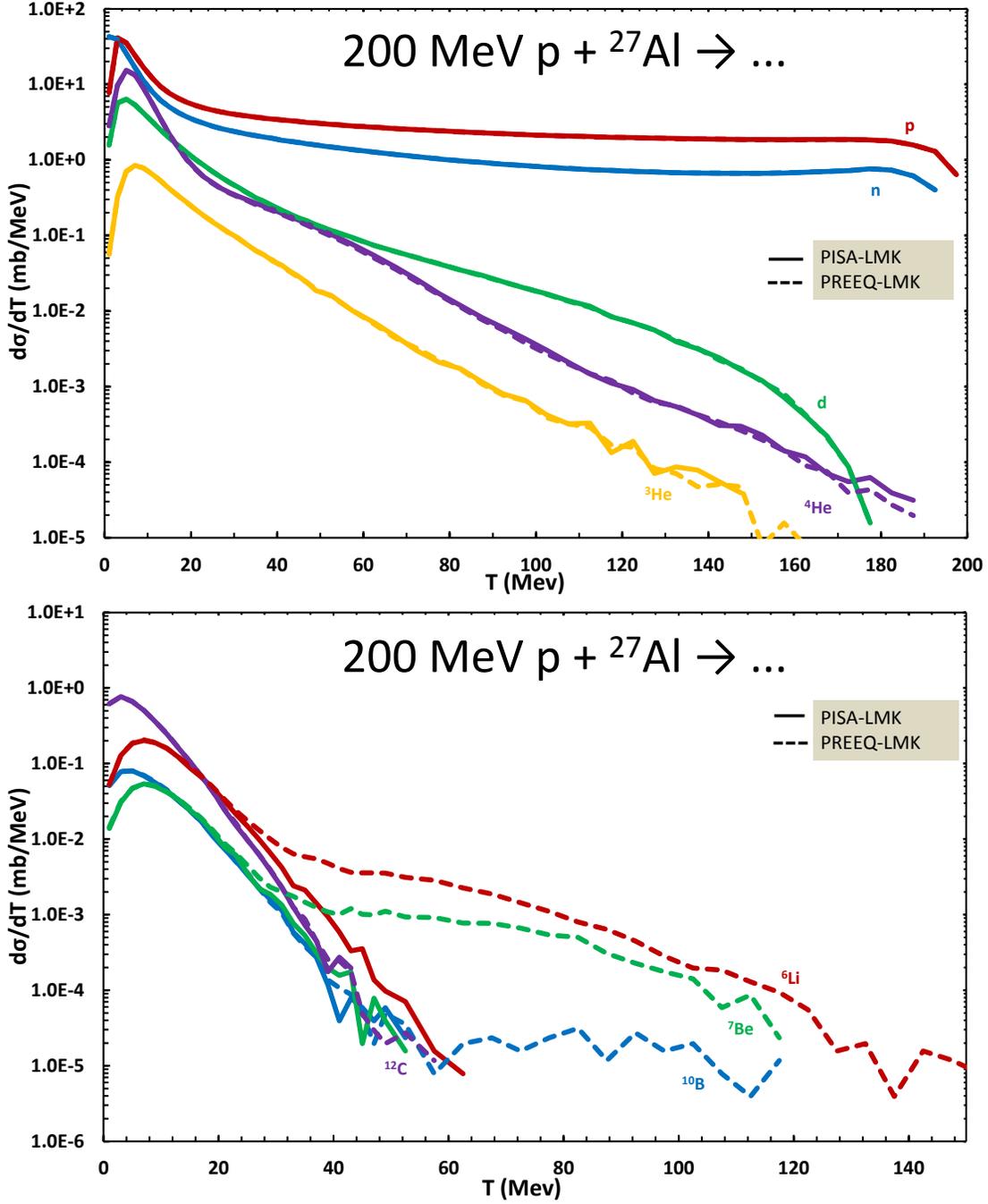}
\vspace*{-2.0in}
\caption[]{Angle integrated cross section using the modified MEM code for the reaction 200 MeV p + $^{27}$Al $\rightarrow$ ...}
\label{fig:PreeqCompAl200Int}
\end{figure}

Figure~\ref{fig:PreeqCompAl200Int} presents energy-spectra of nucleons, d, t, $^3$He, and $^4$He, as well as energy-spectra of heavier fragments $^6$Li, $^7$Be, $^{10}$B, and $^{12}$C. It demonstrates that the modified-MEM code predicts the high-energy tails of light fragment spectra, without destroying the spectra of established particles and fragments.

Table~\ref{p200AlTable} details the $\gamma_\beta$ used in the expanded MEM. At the bottom of the table are values for the residual nuclei energy, E*, atomic number, Z, and mass number, A.

\begin{table}
\caption{$\gamma_\beta$ values for 200 MeV p + $^{27}$Al}
\begin{ruledtabular}
\begin{tabular}{llllllllll}
\hline\hline 
 n & p & d & t & $^3$He & $^4$He & $^6$He & $^8$He & $^6$Li & $^7$Li \\
1.0 & 1.0 & 2.0 & 4.0 & 20.0 & 30.0 & 1.0 & 1.0& 5.0 & 2.0 \\
\hline
$^8$Li & $^{9 }$Li & $^{7 }$Be & $^{9 }$Be &$^{10}$Be & $^{11}$Be & $^{12}$Be & $^{8 }$B & $^{10}$B & $^{11}$B \\
1.7 & 10.0 & 0.3 & 0.2 & 0.2 & 0.2 & 0.2 & 0.2 & 0.2 & 0.1 \\
\hline
$^{12}$B &$^{13}$B &$^{10}$C &$^{11}$C &$^{12}$C &$^{13}$C &$^{14}$C &$^{15}$C &$^{16}$C &$^{12}$N \\
0.1 & 0.1 & 0.1 & 0.1 & 0.1 & 0.1 & 0.1 & 0.1 & 0.1 & 0.1 \\
\hline
$^{13}$N &$^{14}$N &$^{15}$N &$^{16}$N &$^{17}$N &$^{14}$O &$^{15}$O &$^{16}$O &$^{17}$O &$^{18}$O \\
0.1 & 0.1 & 0.1 & 0.1 & 0.1 & 0.1 & 0.1 & 0.1 & 0.1 & 0.1 \\
\hline
$^{19}$O &$^{20}$O &$^{17}$F &$^{18}$F &$^{19}$F &$^{20}$F &$^{21}$F &$^{18}$Ne &$^{19}$Ne &$^{20}$Ne \\
0.1 & 0.1 & 0.1 & 0.1 & 0.1 & 0.1 & 0.1 & 0.1 & 0.1 & 0.1 \\
\hline
$^{21}$Ne &$^{22}$Ne &$^{23}$Ne &$^{24}$Ne &$^{21}$Na &$^{22}$Na &$^{23}$Na &$^{24}$Na &$^{25}$Na &$^{22}$Mg \\
0.1 & 0.1 & 0.1 & 0.1 & 0.1 & 0.1 & 0.1 & 0.1 & 0.1 & 0.1 \\
\hline
$^{23}$Mg &$^{24}$Mg &$^{25}$Mg &$^{26}$Mg &$^{27}$Mg &$^{28}$Mg \\
0.1 & 0.1 & 0.1 & 0.1 & 0.1 & 0.1 \\
\hline\hline 
\multicolumn{4}{l} {E* = 35.0 $\pm$ 33.5 MeV} & \multicolumn{3}{l} {Z = 12.5 $\pm$ 0.8} & \multicolumn{3}{l} {A = 25.9 $\pm$ 0.9} \\
\hline\hline
\end{tabular}
\end{ruledtabular}
\label{p200AlTable}
\end{table}

\subsubsection{190 MeV p + $^{nat}$Ag}
\begin{figure}[]
\centering
\includegraphics[trim = 0.4in 1.5in 0.5in 1.5in, width=6.5in]{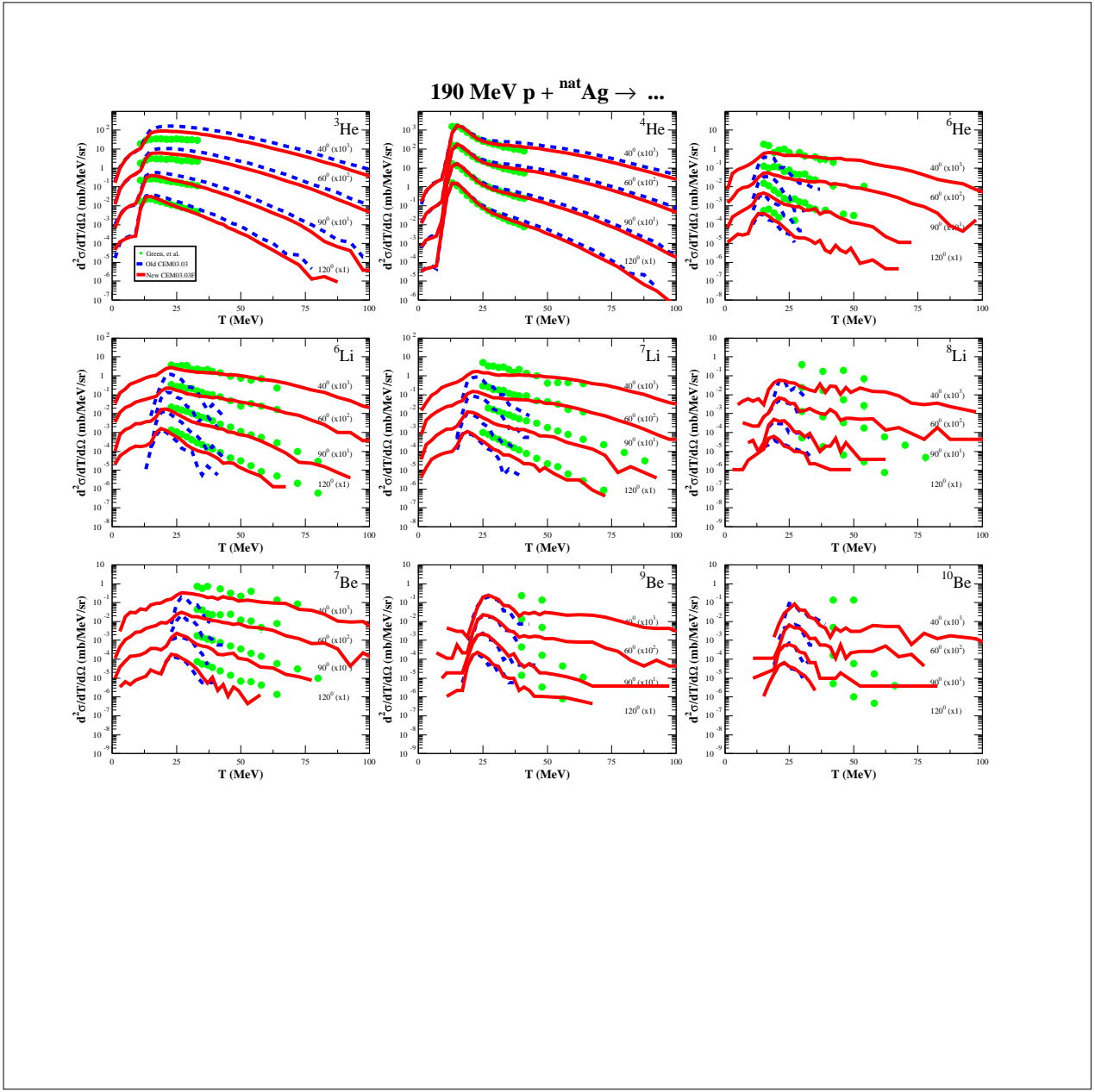}
\caption[]{Comparison of experimental data by Green {\it et al.} \cite{Green} (green points) with results from the unmodified CEM03.03 (blue dotted lines) and the modified MEM CEM03.03 (red solid lines) for 190 MeV p + $^{nat}$Ag $\rightarrow$ $...$}
\label{fig:p190Ag}
\end{figure}

Figure~\ref{fig:p190Ag} demonstrates the potential of the modified precompound code we built for 190 MeV p + $^{nat}$Ag. The red solid lines show results from the new precompound code we designed in FY2013; the blue dotted lines present calculations from the old code; and the green points are experimental data from Green, {\it et al.} \cite{Green}. The upgraded MEM provides dramatically improved ability to describe the cross section at intermediate to high energies. 

Table~\ref{p190AgTable} details the $\gamma_\beta$ used in the expanded MEM. At the bottom of the table are values for the residual nuclei energy, E*, atomic number, Z, and mass number, A.

\begin{table}[]
\caption{$\gamma_\beta$ values for 190 MeV p + $^{nat}$Ag}
\centering
\begin{tabular}{llllllllll}
\hline\hline 
 n & p & d & t & $^3$He & $^4$He & $^6$He & $^8$He & $^6$Li & $^7$Li \\
1.0 & 1.0 & 1.0 & 1.0 & 0.8 & 4.0 & 0.035 & 0.01 & 0.08 & 0.2 \\
\hline
$^8$Li & $^{9 }$Li & $^{7 }$Be & $^{9 }$Be &$^{10}$Be & $^{11}$Be & $^{12}$Be & $^{8 }$B & $^{10}$B & $^{11}$B \\
0.03 & 0.02 & 0.035 & 0.04 & 0.04 & 0.0015 & 0.0015 & 0.0015 & 0.00001 & 0.00001 \\
\hline
$^{12}$B &$^{13}$B &$^{10}$C &$^{11}$C &$^{12}$C &$^{13}$C &$^{14}$C &$^{15}$C &$^{16}$C &$^{12}$N \\
0.0001 & 0.0001 & 0.0001 & 0.0001 & 0.00001 & 0.00001 & 0.00001 & 0.00001 & 0.00001 & 0.00001 \\
\hline
$^{13}$N &$^{14}$N &$^{15}$N &$^{16}$N &$^{17}$N &$^{14}$O &$^{15}$O &$^{16}$O &$^{17}$O &$^{18}$O \\
0.00001 & 0.00001 & 0.00001 & 0.00001 & 0.00001 & 0.00001 & 0.00001 & 0.00001 & 0.00001 & 0.00001 \\
\hline
$^{19}$O &$^{20}$O &$^{17}$F &$^{18}$F &$^{19}$F &$^{20}$F &$^{21}$F &$^{18}$Ne &$^{19}$Ne &$^{20}$Ne \\
0.00001 & 0.00001 & 0.00001 & 0.00001 & 0.00001 & 0.00001 & 0.00001 & 0.00001 & 0.00001 & 0.00001 \\
\hline
$^{21}$Ne &$^{22}$Ne &$^{23}$Ne &$^{24}$Ne &$^{21}$Na &$^{22}$Na &$^{23}$Na &$^{24}$Na &$^{25}$Na &$^{22}$Mg \\
0.00001 & 0.00001 & 0.00001 & 0.00001 & 0.00001 & 0.00001 & 0.00001 & 0.00001 & 0.00001 & 0.00001 \\
\hline
$^{23}$Mg &$^{24}$Mg &$^{25}$Mg &$^{26}$Mg &$^{27}$Mg &$^{28}$Mg \\
0.00001 & 0.00001 & 0.00001 & 0.00001 & 0.00001 & 0.00001 \\
\hline\hline 
\multicolumn{4}{l} {E* = 68.6 $\pm$ 49.1 MeV} & \multicolumn{3}{l} {Z = 47.1 $\pm$ 0.7} & \multicolumn{3}{l} {A = 106.2 $\pm$ 0.5} \\
\hline\hline
\end{tabular}
\label{p190AgTable}
\end{table}

\subsubsection{300 MeV p + $^{nat}$Ag}
\begin{figure}[]
\centering
\includegraphics[trim = 0.4in 1.5in 0.5in 1.5in, width=6.5in]{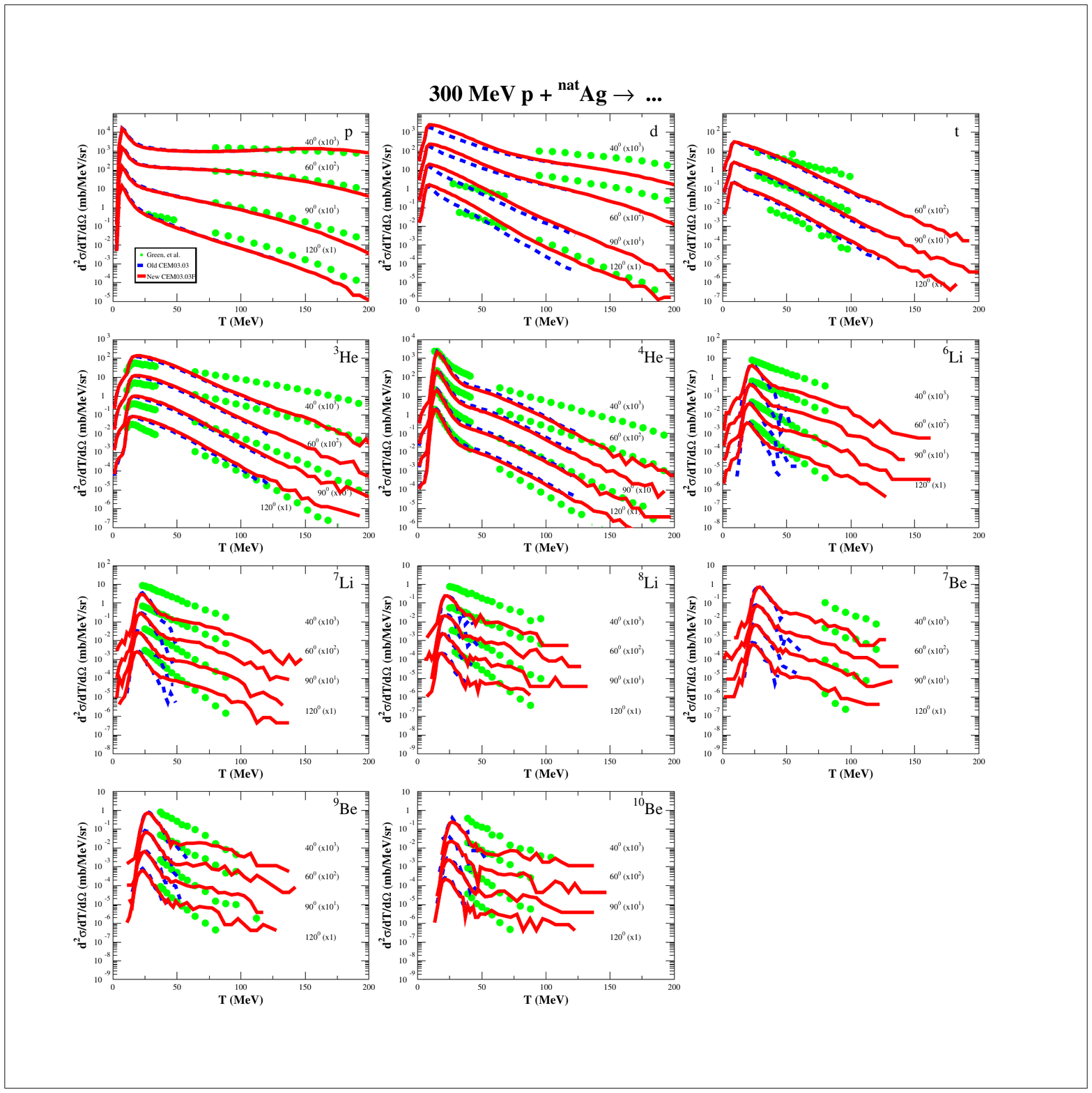}
\caption[]{Comparison of experimental data by Green {\it et al.} \cite{Green} (green points) with results from the unmodified CEM03.03 (blue dotted lines) and the modified MEM CEM03.03 (red solid lines) for 300 MeV p + $^{nat}$Ag $\rightarrow$ $...$}
\label{fig:p300Ag}
\end{figure}

Figure~\ref{fig:p300Ag} demonstrates the potential of the modified precompound code we built for 300 MeV p + $^{nat}$Ag. The red solid lines show results from the new precompound code we designed in FY2013; the blue dotted lines present calculations from the old code; and the green points are experimental data from Green, {\it et al.} \cite{Green}. The upgraded MEM provides dramatically improved ability to describe the cross section at intermediate to high energies. 

Table~\ref{p300AgTable} details the $\gamma_\beta$ used in the expanded MEM. At the bottom of the table are values for the residual nuclei energy, E*, atomic number, Z, and mass number, A.

\begin{table}[]
\caption{$\gamma_\beta$ values for 300 MeV p + $^{nat}$Ag}
\centering
\begin{tabular}{llllllllll}
\hline\hline 
 n & p & d & t & $^3$He & $^4$He & $^6$He & $^8$He & $^6$Li & $^7$Li \\
1.0 & 1.0 & 6.0 & 3.0 & 3.0 & 2.0 & 0.01 & 0.01 & 0.012 & 0.01 \\
\hline
$^8$Li & $^{9 }$Li & $^{7 }$Be & $^{9 }$Be &$^{10}$Be & $^{11}$Be & $^{12}$Be & $^{8 }$B & $^{10}$B & $^{11}$B \\
0.0025 & 0.0022 & 0.002 & 0.002 & 0.0015 & 0.0015 & 0.0015 & 0.0015 & 0.00001 & 0.00001 \\
\hline
$^{12}$B &$^{13}$B &$^{10}$C &$^{11}$C &$^{12}$C &$^{13}$C &$^{14}$C &$^{15}$C &$^{16}$C &$^{12}$N \\
0.0001 & 0.0001 & 0.0001 & 0.0001 & 0.00001 & 0.00001 & 0.00001 & 0.00001 & 0.00001 & 0.00001 \\
\hline
$^{13}$N &$^{14}$N &$^{15}$N &$^{16}$N &$^{17}$N &$^{14}$O &$^{15}$O &$^{16}$O &$^{17}$O &$^{18}$O \\
0.00001 & 0.00001 & 0.00001 & 0.00001 & 0.00001 & 0.00001 & 0.00001 & 0.00001 & 0.00001 & 0.00001 \\
\hline
$^{19}$O &$^{20}$O &$^{17}$F &$^{18}$F &$^{19}$F &$^{20}$F &$^{21}$F &$^{18}$Ne &$^{19}$Ne &$^{20}$Ne \\
0.00001 & 0.00001 & 0.00001 & 0.00001 & 0.00001 & 0.00001 & 0.00001 & 0.00001 & 0.00001 & 0.00001 \\
\hline
$^{21}$Ne &$^{22}$Ne &$^{23}$Ne &$^{24}$Ne &$^{21}$Na &$^{22}$Na &$^{23}$Na &$^{24}$Na &$^{25}$Na &$^{22}$Mg \\
0.00001 & 0.00001 & 0.00001 & 0.00001 & 0.00001 & 0.00001 & 0.00001 & 0.00001 & 0.00001 & 0.00001 \\
\hline
$^{23}$Mg &$^{24}$Mg &$^{25}$Mg &$^{26}$Mg &$^{27}$Mg &$^{28}$Mg \\
0.00001 & 0.00001 & 0.00001 & 0.00001 & 0.00001 & 0.00001 \\
\hline\hline 
\multicolumn{4}{l} {E* = 83.4 $\pm$ 63.4 MeV} & \multicolumn{3}{l} {Z = 46.8 $\pm$ 0.8} & \multicolumn{3}{l} {A = 105.3 $\pm$ 1.2} \\
\hline\hline
\end{tabular}
\label{p300AgTable}
\end{table}

\subsubsection{480 MeV p + $^{nat}$Ag}
\begin{figure}[]
\centering
\includegraphics[trim = 0.4in 1.5in 0.5in 1.5in, width=6.5in]{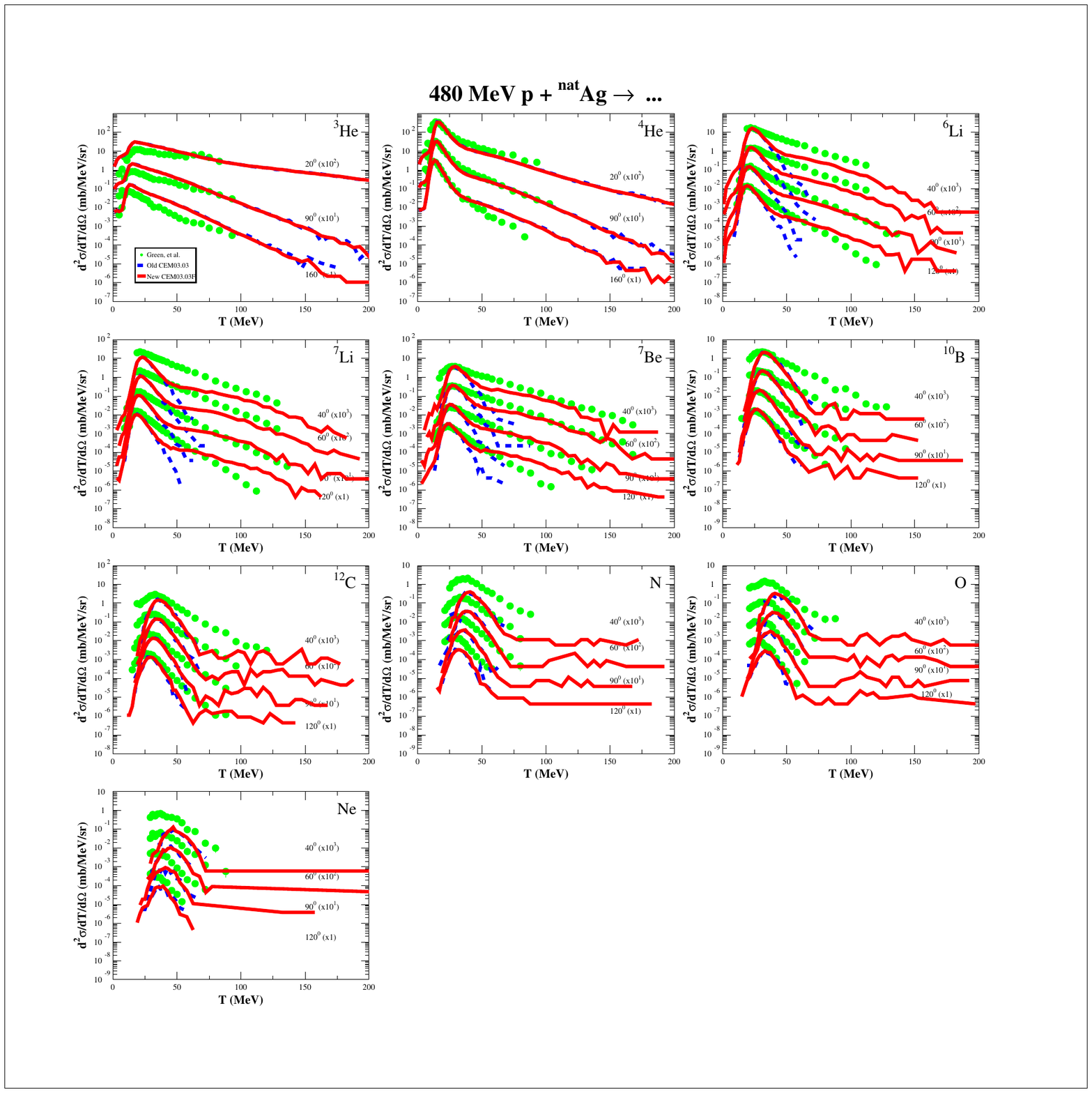}
\caption[]{Comparison of experimental data by Green {\it et al.} \cite{Green480} (green points) with results from the unmodified CEM03.03 (blue dotted lines) and the modified MEM CEM03.03 (red solid lines) for 480 MeV p + $^{nat}$Ag $\rightarrow$ $...$}
\label{fig:p480Ag}
\end{figure}

Figure~\ref{fig:p480Ag} demonstrates the potential of the modified precompound code we built for 480 MeV p + $^{nat}$Ag. The red solid lines show results from the new precompound code we designed in FY2013; the blue dotted lines present calculations from the old code; and the green points are experimental data from Green, {\it et al.} \cite{Green480}. The upgraded MEM provides dramatically improved ability to describe the cross section at intermediate to high energies. 

Table~\ref{p480AgTable} details the $\gamma_\beta$ used in the expanded MEM. At the bottom of the table are values for the residual nuclei energy, E*, atomic number, Z, and mass number, A.

\begin{table}[]
\caption{$\gamma_\beta$ values for 480 MeV p + $^{nat}$Ag}
\centering
\begin{tabular}{llllllllll}
\hline\hline 
 n & p & d & t & $^3$He & $^4$He & $^6$He & $^8$He & $^6$Li & $^7$Li \\
1.0 & 1.0 & 1.0 & 1.0 & 1.0 & 1.0 & 0.01 & 0.01 & 0.002 & 0.001 \\
\hline
$^8$Li & $^{9 }$Li & $^{7 }$Be & $^{9 }$Be &$^{10}$Be & $^{11}$Be & $^{12}$Be & $^{8 }$B & $^{10}$B & $^{11}$B \\
0.0022 & 0.0022 & 0.0008 & 0.0015 & 0.0015 & 0.0015 & 0.0015 & 0.0015 & 0.00001 & 0.00001 \\
\hline
$^{12}$B &$^{13}$B &$^{10}$C &$^{11}$C &$^{12}$C &$^{13}$C &$^{14}$C &$^{15}$C &$^{16}$C &$^{12}$N \\
0.0001 & 0.0001 & 0.0001 & 0.0001 & 0.00001 & 0.00001 & 0.00001 & 0.00001 & 0.00001 & 0.00001 \\
\hline
$^{13}$N &$^{14}$N &$^{15}$N &$^{16}$N &$^{17}$N &$^{14}$O &$^{15}$O &$^{16}$O &$^{17}$O &$^{18}$O \\
0.00001 & 0.00001 & 0.00001 & 0.00001 & 0.00001 & 0.00001 & 0.00001 & 0.00001 & 0.00001 & 0.00001 \\
\hline
$^{19}$O &$^{20}$O &$^{17}$F &$^{18}$F &$^{19}$F &$^{20}$F &$^{21}$F &$^{18}$Ne &$^{19}$Ne &$^{20}$Ne \\
0.00001 & 0.00001 & 0.00001 & 0.00001 & 0.00001 & 0.00001 & 0.00001 & 0.00001 & 0.00001 & 0.00001 \\
\hline
$^{21}$Ne &$^{22}$Ne &$^{23}$Ne &$^{24}$Ne &$^{21}$Na &$^{22}$Na &$^{23}$Na &$^{24}$Na &$^{25}$Na &$^{22}$Mg \\
0.00001 & 0.00001 & 0.00001 & 0.00001 & 0.00001 & 0.00001 & 0.00001 & 0.00001 & 0.00001 & 0.00001 \\
\hline
$^{23}$Mg &$^{24}$Mg &$^{25}$Mg &$^{26}$Mg &$^{27}$Mg &$^{28}$Mg \\
0.00001 & 0.00001 & 0.00001 & 0.00001 & 0.00001 & 0.00001 \\
\hline\hline 
\multicolumn{4}{l} {E* = 113.4 $\pm$ 87.6 MeV} & \multicolumn{3}{l} {Z = 46.6 $\pm$ 1.0} & \multicolumn{3}{l} {A = 104.6 $\pm$ 1.7} \\
\hline\hline
\end{tabular}
\label{p480AgTable}
\end{table}

\subsubsection{1200 MeV p + $^{197}$Au}
\begin{figure}[]
\centering
\includegraphics[trim = 0.5in 1.5in 0.5in 1.5in, width=6.5in]{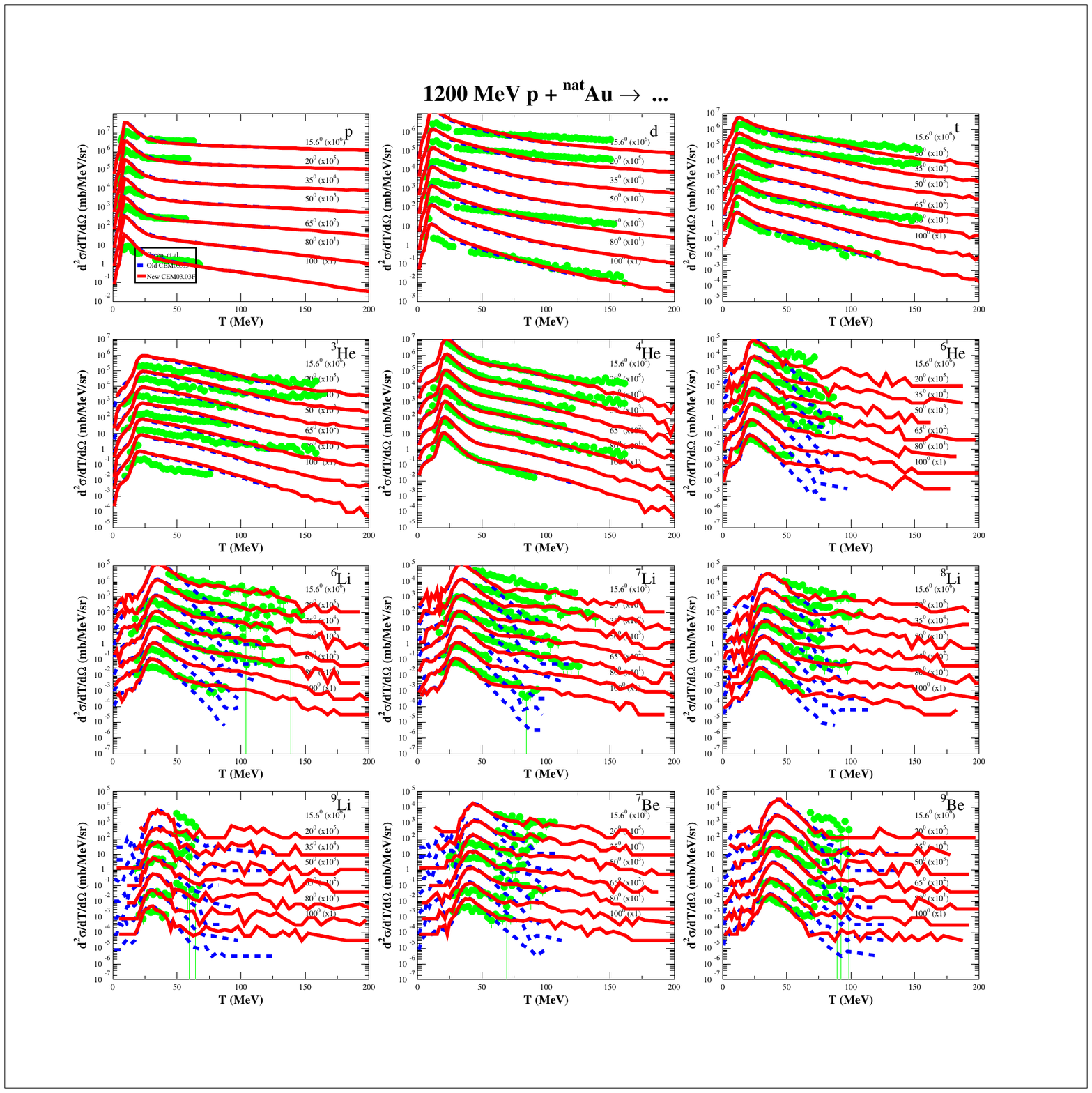}
\caption[]{Comparison of experimental data by Budzanowski, {\it et al.} \cite{Budzanowski} (green points) with results from the unmodified CEM03.03 (blue dashed lines) and the modified-MEM CEM03.03 (red solid lines) for 1200 MeV p + $^{197}$Au $\rightarrow$ $...$}
\label{p1200Au}
\end{figure}

We also have good preliminary results for a reaction where a higher-energy incident particle strikes a larger target mass. Figure~\ref{p1200Au} compares experimental data by Budzanowski, {\it et al.} \cite{Budzanowski} with results by the unmodified CEM03.03 and the modified-MEM CEM03.03  for the reaction 1200 MeV p + $^{197}$Au $\rightarrow$ $...$.

Table~\ref{p1200AuTable} details the $\gamma_\beta$ used in the expanded MEM. At the bottom of the table are values for the residual nuclei energy, E*, atomic number, Z, and mass number, A.

\begin{table}[]
\caption{$\gamma_\beta$ values for 1200 MeV p + $^{197}$Au}
\centering
\begin{tabular}{llllllllll}
\hline\hline 
 n & p & d & t & $^3$He & $^4$He & $^6$He & $^8$He & $^6$Li & $^7$Li \\
1.0 & 1.0 & 2.0 & 4.0 & 1.5 & 2.0 & 0.0004 & 0.0012 & 0.001 & 0.00007 \\
\hline
$^8$Li & $^{9 }$Li & $^{7 }$Be & $^{9 }$Be &$^{10}$Be & $^{11}$Be & $^{12}$Be & $^{8 }$B & $^{10}$B & $^{11}$B \\
0.00007 & 0.00001 & 0.00015 & 0.00001 & 0.0002 & 0.0002 & 0.0002 & 0.0002 & 0.00001 & 0.00001 \\
\hline
$^{12}$B &$^{13}$B &$^{10}$C &$^{11}$C &$^{12}$C &$^{13}$C &$^{14}$C &$^{15}$C &$^{16}$C &$^{12}$N \\
0.0001 & 0.0001 & 0.0001 & 0.0001 & 0.00001 & 0.00001 & 0.00001 & 0.00001 & 0.00001 & 0.00001 \\
\hline
$^{13}$N &$^{14}$N &$^{15}$N &$^{16}$N &$^{17}$N &$^{14}$O &$^{15}$O &$^{16}$O &$^{17}$O &$^{18}$O \\
0.00001 & 0.00001 & 0.00001 & 0.00001 & 0.00001 & 0.00001 & 0.00001 & 0.00001 & 0.00001 & 0.00001 \\
\hline
$^{19}$O &$^{20}$O &$^{17}$F &$^{18}$F &$^{19}$F &$^{20}$F &$^{21}$F &$^{18}$Ne &$^{19}$Ne &$^{20}$Ne \\
0.00001 & 0.00001 & 0.00001 & 0.00001 & 0.00001 & 0.00001 & 0.00001 & 0.00001 & 0.00001 & 0.00001 \\
\hline
$^{21}$Ne &$^{22}$Ne &$^{23}$Ne &$^{24}$Ne &$^{21}$Na &$^{22}$Na &$^{23}$Na &$^{24}$Na &$^{25}$Na &$^{22}$Mg \\
0.00001 & 0.00001 & 0.00001 & 0.00001 & 0.00001 & 0.00001 & 0.00001 & 0.00001 & 0.00001 & 0.00001 \\
\hline
$^{23}$Mg &$^{24}$Mg &$^{25}$Mg &$^{26}$Mg &$^{27}$Mg &$^{28}$Mg \\
0.00001 & 0.00001 & 0.00001 & 0.00001 & 0.00001 & 0.00001 \\
\hline\hline 
\multicolumn{4}{l} {E* = 320.6 $\pm$ 208.0 MeV} & \multicolumn{3}{l} {Z = 78.2 $\pm$ 1.3} & \multicolumn{3}{l} {A = 190.9 $\pm$ 3.7} \\
\hline\hline
\end{tabular}
\label{p1200AuTable}
\end{table}

The modified-MEM CEM03.03 peak of spectra at low energies exceeds the peak of the experimental data. However, this is an issue with the evaporative stage and our work has thus far focused on the precompound stages. The high-energy tails do match the experimental data well with this new modified MEM code. The evaporation model can be fixed, and we would like to do so in the future.

\subsubsection{1200 MeV p + $^{nat}$Ni}
\begin{figure}[]
\centering
\includegraphics[trim = 0.5in 1.5in 0.5in 1.5in, width=6.5in]{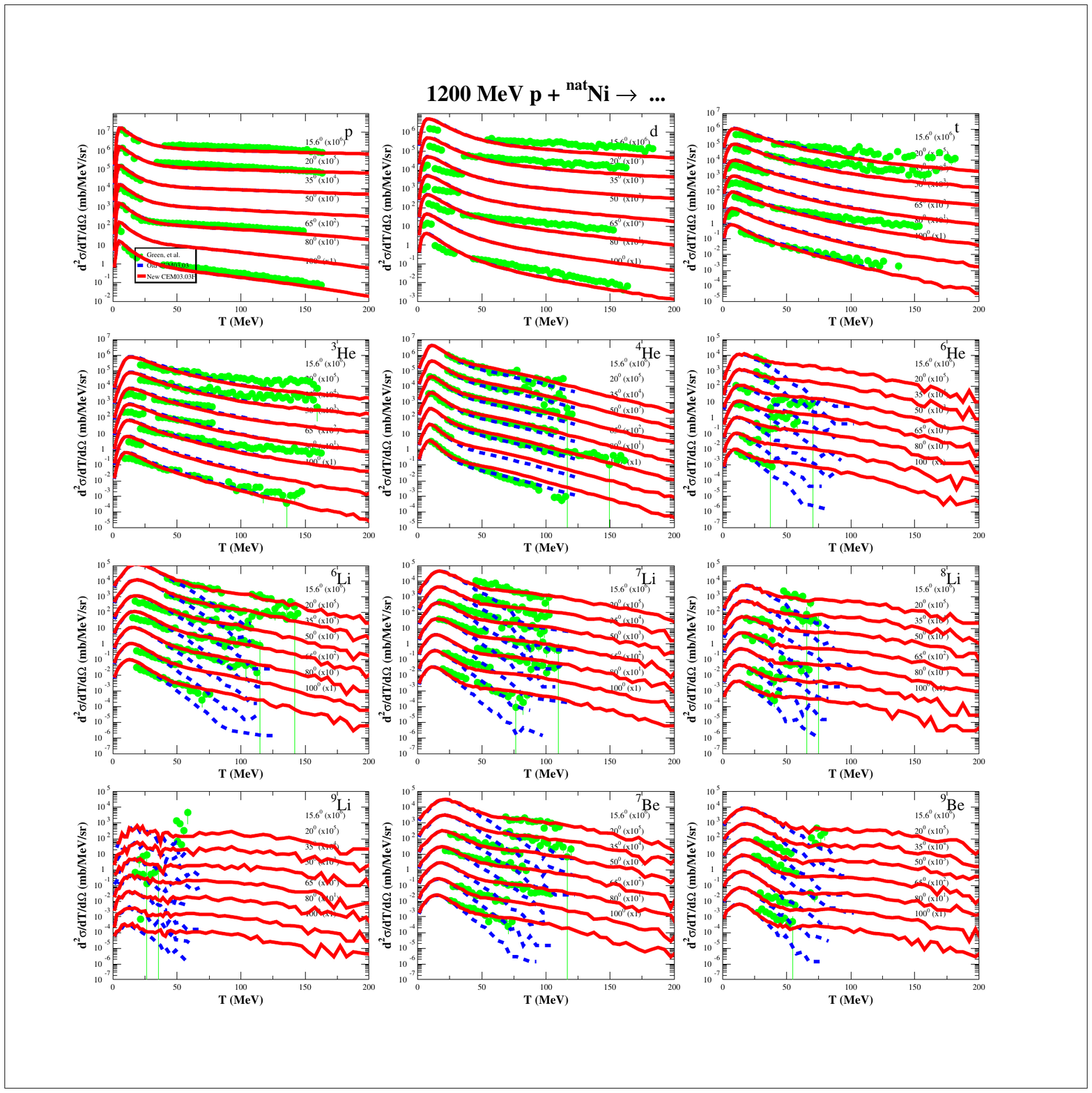}
\caption[]{Comparison of experimental data by Budzanowski, {\it et al.} \cite{BudzanowskiNi} (green points) with results from the unmodified CEM03.03 (blue dashed lines) and the modified-MEM CEM03.03 (red solid lines) for 1200 MeV p + $^{nat}$Ni $\rightarrow$ $...$}
\label{p1200Ni}
\end{figure}

Figure~\ref{p1200Ni} compares experimental data by Budzanowski, {\it et al.} \cite{BudzanowskiNi} with results by the unmodified CEM03.03 and the modified-MEM CEM03.03  for the reaction 1200 MeV p + $^{nat}$Ni $\rightarrow$ $...$.

Table~\ref{p1200NiTable} details the $\gamma_\beta$ used in the expanded MEM. At the bottom of the table are values for the residual nuclei energy, E*, atomic number, Z, and mass number, A.

\begin{table}[]
\caption{$\gamma_\beta$ values for 1200 MeV p + $^{61}$Ni}
\centering
\begin{tabular}{llllllllll}
\hline\hline 
 n & p & d & t & $^3$He & $^4$He & $^6$He & $^8$He & $^6$Li & $^7$Li \\
1.0 & 1.0 & 1.0 & 1.0 & 1.0 & 4.0 & 0.008 & 0.01 & 0.007 & 0.003 \\
\hline
$^8$Li & $^{9 }$Li & $^{7 }$Be & $^{9 }$Be &$^{10}$Be & $^{11}$Be & $^{12}$Be & $^{8 }$B & $^{10}$B & $^{11}$B \\
0.001 & 0.0004 & 0.002 & 0.0002 & 0.0002 & 0.0002 & 0.0002 & 0.0002 & 0.0001 & 0.0001 \\
\hline
$^{12}$B &$^{13}$B &$^{10}$C &$^{11}$C &$^{12}$C &$^{13}$C &$^{14}$C &$^{15}$C &$^{16}$C &$^{12}$N \\
0.0001 & 0.0001 & 0.0001 & 0.0001 & 0.0001 & 0.0001 & 0.0001 & 0.0001 & 0.0001 & 0.0001 \\
\hline
$^{13}$N &$^{14}$N &$^{15}$N &$^{16}$N &$^{17}$N &$^{14}$O &$^{15}$O &$^{16}$O &$^{17}$O &$^{18}$O \\
0.0001 & 0.0001 & 0.0001 & 0.0001 & 0.0001 & 0.0001 & 0.0001 & 0.0001 & 0.0001 & 0.0001 \\
\hline
$^{19}$O &$^{20}$O &$^{17}$F &$^{18}$F &$^{19}$F &$^{20}$F &$^{21}$F &$^{18}$Ne &$^{19}$Ne &$^{20}$Ne \\
0.0001 & 0.0001 & 0.0001 & 0.0001 & 0.0001 & 0.0001 & 0.0001 & 0.0001 & 0.0001 & 0.0001 \\
\hline
$^{21}$Ne &$^{22}$Ne &$^{23}$Ne &$^{24}$Ne &$^{21}$Na &$^{22}$Na &$^{23}$Na &$^{24}$Na &$^{25}$Na &$^{22}$Mg \\
0.0001 & 0.0001 & 0.0001 & 0.0001 & 0.0001 & 0.0001 & 0.0001 & 0.0001 & 0.0001 & 0.0001 \\
\hline
$^{23}$Mg &$^{24}$Mg &$^{25}$Mg &$^{26}$Mg &$^{27}$Mg &$^{28}$Mg \\
0.0001 & 0.0001 & 0.0001 & 0.0001 & 0.0001 & 0.0001 \\
\hline\hline 
\multicolumn{4}{l} {E* = 177.2 $\pm$ 140.3 MeV} & \multicolumn{3}{l} {Z = 26.5 $\pm$ 1.6} & \multicolumn{3}{l} {A = 56.4 $\pm$ 3.3} \\
\hline\hline
\end{tabular}
\label{p1200NiTable}
\end{table}

\subsubsection{1900 MeV p + $^{nat}$Ni}
\begin{figure}[]
\centering
\includegraphics[trim = 0.5in 1.5in 0.5in 1.5in, width=6.5in]{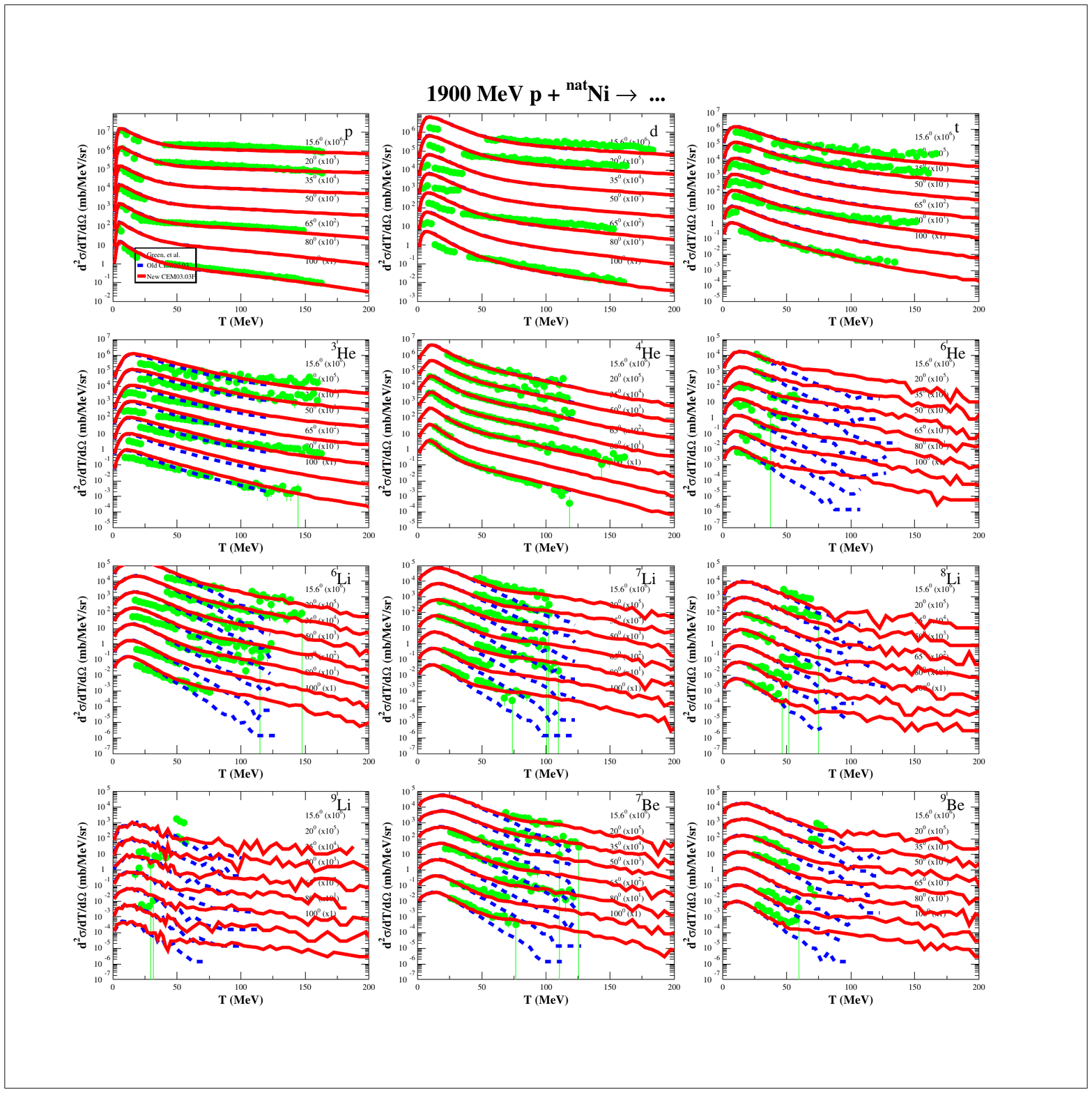}
\caption[]{Comparison of experimental data by Budzanowski, {\it et al.} \cite{BudzanowskiNi} (green points) with results from the unmodified CEM03.03 (blue dashed lines) and the modified-MEM CEM03.03 (red solid lines) for 1900 MeV p + $^{nat}$Ni $\rightarrow$ $...$}
\label{p1900Ni}
\end{figure}

Figure~\ref{p1900Ni} compares experimental data by Budzanowski, {\it et al.} \cite{BudzanowskiNi} with results by the unmodified CEM03.03 and the modified-MEM CEM03.03  for the reaction 1900 MeV p + $^{nat}$Ni $\rightarrow$ $...$.

Table~\ref{p1900NiTable} details the $\gamma_\beta$ used in the expanded MEM. At the bottom of the table are values for the residual nuclei energy, E*, atomic number, Z, and mass number, A.

\begin{table}[]
\caption{$\gamma_\beta$ values for 1900 MeV p + $^{61}$Ni}
\centering
\begin{tabular}{llllllllll}
\hline\hline 
 n & p & d & t & $^3$He & $^4$He & $^6$He & $^8$He & $^6$Li & $^7$Li \\
1.0 & 1.0 & 1.5 & 2.0 & 8.0 & 4.0 & 0.004 & 0.01 & 0.007 & 0.002 \\
\hline
$^8$Li & $^{9 }$Li & $^{7 }$Be & $^{9 }$Be &$^{10}$Be & $^{11}$Be & $^{12}$Be & $^{8 }$B & $^{10}$B & $^{11}$B \\
0.0001 & 0.00005 & 0.001 & 0.00005 & 0.0001 & 0.0001 & 0.0001 & 0.0001 & 0.0001 & 0.0001 \\
\hline
$^{12}$B &$^{13}$B &$^{10}$C &$^{11}$C &$^{12}$C &$^{13}$C &$^{14}$C &$^{15}$C &$^{16}$C &$^{12}$N \\
0.0001 & 0.0001 & 0.0001 & 0.0001 & 0.0001 & 0.0001 & 0.0001 & 0.0001 & 0.0001 & 0.0001 \\
\hline
$^{13}$N &$^{14}$N &$^{15}$N &$^{16}$N &$^{17}$N &$^{14}$O &$^{15}$O &$^{16}$O &$^{17}$O &$^{18}$O \\
0.0001 & 0.0001 & 0.0001 & 0.0001 & 0.0001 & 0.0001 & 0.0001 & 0.0001 & 0.0001 & 0.0001 \\
\hline
$^{19}$O &$^{20}$O &$^{17}$F &$^{18}$F &$^{19}$F &$^{20}$F &$^{21}$F &$^{18}$Ne &$^{19}$Ne &$^{20}$Ne \\
0.0001 & 0.0001 & 0.0001 & 0.0001 & 0.0001 & 0.0001 & 0.0001 & 0.0001 & 0.0001 & 0.0001 \\
\hline
$^{21}$Ne &$^{22}$Ne &$^{23}$Ne &$^{24}$Ne &$^{21}$Na &$^{22}$Na &$^{23}$Na &$^{24}$Na &$^{25}$Na &$^{22}$Mg \\
0.0001 & 0.0001 & 0.0001 & 0.0001 & 0.0001 & 0.0001 & 0.0001 & 0.0001 & 0.0001 & 0.0001 \\
\hline
$^{23}$Mg &$^{24}$Mg &$^{25}$Mg &$^{26}$Mg &$^{27}$Mg &$^{28}$Mg \\
0.0001 & 0.0001 & 0.0001 & 0.0001 & 0.0001 & 0.0001 \\
\hline\hline 
\multicolumn{4}{l} {E* = 242.7 $\pm$ 194.9 MeV} & \multicolumn{3}{l} {Z = 25.8 $\pm$ 2.1} & \multicolumn{3}{l} {A = 54.7 $\pm$ 4.6} \\
\hline\hline
\end{tabular}
\label{p1900NiTable}
\end{table}

\subsubsection{2500 MeV p + $^{nat}$Ni}
\begin{figure}[]
\centering
\includegraphics[trim = 0.5in 1.5in 0.5in 1.5in, width=6.5in]{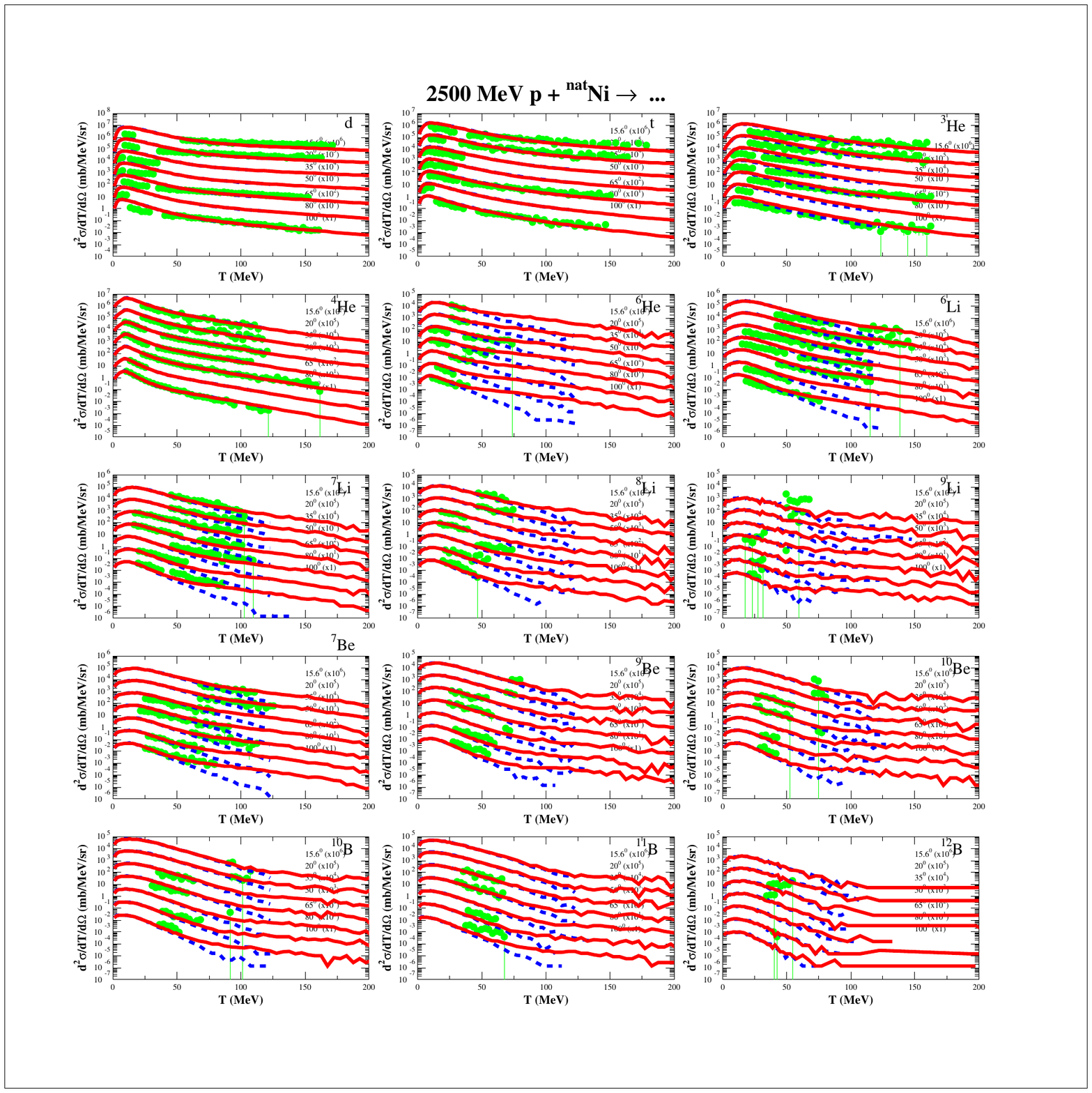}
\caption[]{Comparison of experimental data by Budzanowski, {\it et al.} \cite{BudzanowskiNi} (green points) with results from the unmodified CEM03.03 (blue dashed lines) and the modified-MEM CEM03.03 (red solid lines) for 2500 MeV p + $^{nat}$Ni $\rightarrow$ $...$}
\label{p2500Ni}
\end{figure}

Figure~\ref{p2500Ni} compares experimental data by Budzanowskik, {\it et al.} \cite{BudzanowskiNi} with results by the unmodified CEM03.03 and the modified-MEM CEM03.03  for the reaction 2500 MeV p + $^{nat}$Ni $\rightarrow$ $...$.

Table~\ref{p2500NiTable} details the $\gamma_\beta$ used in the expanded MEM. At the bottom of the table are values for the residual nuclei energy, E*, atomic number, Z, and mass number, A.

\begin{table}[]
\caption{$\gamma_\beta$ values for 2500 MeV p + $^{61}$Ni}
\centering
\begin{tabular}{llllllllll}
\hline\hline 
 n & p & d & t & $^3$He & $^4$He & $^6$He & $^8$He & $^6$Li & $^7$Li \\
1.0 & 1.0 & 1.5 & 2.0 & 7.0 & 4.0 & 0.004 & 0.01 & 0.007 & 0.0007 \\
\hline
$^8$Li & $^{9 }$Li & $^{7 }$Be & $^{9 }$Be &$^{10}$Be & $^{11}$Be & $^{12}$Be & $^{8 }$B & $^{10}$B & $^{11}$B \\
0.00007 & 0.00003 & 0.001 & 0.000015 & 0.00001 & 0.00003 & 0.00003 & 0.00003 & 0.00001 & 0.000004 \\
\hline
$^{12}$B &$^{13}$B &$^{10}$C &$^{11}$C &$^{12}$C &$^{13}$C &$^{14}$C &$^{15}$C &$^{16}$C &$^{12}$N \\
0.0000001 & 0.0001 & 0.0001 & 0.0001 & 0.0001 & 0.0001 & 0.0001 & 0.0001 & 0.0001 & 0.0001 \\
\hline
$^{13}$N &$^{14}$N &$^{15}$N &$^{16}$N &$^{17}$N &$^{14}$O &$^{15}$O &$^{16}$O &$^{17}$O &$^{18}$O \\
0.0001 & 0.0001 & 0.0001 & 0.0001 & 0.0001 & 0.0001 & 0.0001 & 0.0001 & 0.0001 & 0.0001 \\
\hline
$^{19}$O &$^{20}$O &$^{17}$F &$^{18}$F &$^{19}$F &$^{20}$F &$^{21}$F &$^{18}$Ne &$^{19}$Ne &$^{20}$Ne \\
0.0001 & 0.0001 & 0.0001 & 0.0001 & 0.0001 & 0.0001 & 0.0001 & 0.0001 & 0.0001 & 0.0001 \\
\hline
$^{21}$Ne &$^{22}$Ne &$^{23}$Ne &$^{24}$Ne &$^{21}$Na &$^{22}$Na &$^{23}$Na &$^{24}$Na &$^{25}$Na &$^{22}$Mg \\
0.0001 & 0.0001 & 0.0001 & 0.0001 & 0.0001 & 0.0001 & 0.0001 & 0.0001 & 0.0001 & 0.0001 \\
\hline
$^{23}$Mg &$^{24}$Mg &$^{25}$Mg &$^{26}$Mg &$^{27}$Mg &$^{28}$Mg \\
0.0001 & 0.0001 & 0.0001 & 0.0001 & 0.0001 & 0.0001 \\
\hline\hline 
\multicolumn{4}{l} {E* = 296.8 $\pm$ 236.6 MeV} & \multicolumn{3}{l} {Z = 25.3 $\pm$ 2.4} & \multicolumn{3}{l} {A = 53.3 $\pm$ 5.7} \\
\hline\hline
\end{tabular}
\label{p2500NiTable}
\end{table}

\section{Future Work}
Preliminary results are very encouraging, but we have more work to do to generalize our new MEM across all reactions and a greater spectrum of possible light fragments, and we would like to investigate other modes of precompound emission of high-energy light fragments such as coalescence and Fermi break-up. Future work we plan to undertake:
\begin{enumerate}
\item{Complete the parameterization of $\gamma_j$ in Eq.~\ref{GammaBeta};}
\item{Perform comparisons with the modified-MEM CEM03.03 and Hagiwara's \cite{Hagiwara} data for heavier silicon and aluminum targets, as well as with other reactions we find reliable data for;}
\item{Implement the MEM upgrades in LAQGSM;}
\item{Upgrade the GENXS particle tallies within MCNP6 so that output can be obtained for LF spectra beyond $^4$He;}
\item{Expand the coalescence model to also emit light fragments heavier than $^4$He. This will enable us to describe higher-energy cross sections of light fragments, even beyond what the new MEM can now;}
\item{Upgrade our Fermi break-up model to more physically emit light fragments.}
\end{enumerate}

\section{Conclusion}
We successfully finished our expansion of the MEM to include 66 particles (light fragments up to $^{28}$Mg). Our previous work the summer of 2012 had only expanded the MEM to 26 particles (up to $^{14}$C). We also implemented code crash protection throughout the entirety of CEM. Lastly, we have begun the tedious work of recalibrating MEM parameters across a broad range of reactions.

Our results demonstrate that modifying CEM to simulate precompound emission of light fragments yields better cross sections of intermediate- and high-energy light fragments. Comparisons with several reactions, including experimental results obtained by Machner {\it et al.} \cite{Machner}, Bubak {\it et al.} \cite{Bubak}, and Budzanowski {\it et al.} \cite{Budzanowski}, demonstrate the potential of the new MEM we built to correctly predict high-energy spectra of light fragments. 

Our preliminary results indicate our new MEM works well across different energy regimes, for both light and heavy targets. However, more work is necessary to generalize the new MEM across arbitrary reactions.

\clearpage

\section{Acknowledgements}
One of us (LMK) is grateful to
\begin{enumerate}[label=\emph{\alph*})]
\item{Dr. Stepan Mashnik, for his continued mentoring and ample technical and scientific support and encouragement;}
\item{Dr. Tim Goorley and Los Alamos National Laboratory for the opportunity to study with some of the world's greatest experts in nuclear physics, particularly high-energy physics.}
\item{Dr. Akira Tokuhiro, for his continued support and expertise in serving as my thesis advisor;}
\end{enumerate}

This study was carried out under the auspices of the National Nuclear Security Administration of the U.S. Department of Energy at Los Alamos National Laboratory under Contract No. DE-AC52-06NA253996.

\bibliographystyle{report}
\bibliography{FY2013ReportReferences}

\newpage
\appendix

\section{Distributions of Residual Nuclei After INC}
To understand the mechanisms of nuclear reactions better, we need to have information about various physical properties of our residual nuclei (such as momentum, angular momentum, energy, A and Z numbers, and exciton information) at various stages of the spallation reaction. We therefore built a module to calculate and output these residual nuclei physical properties. The module can be inserted anywhere in the reaction process we want to investigate.

\begin{figure}[htp]
\centering
\includegraphics[trim = 0in 1.0in 0in 1.0in, width=5.0in]{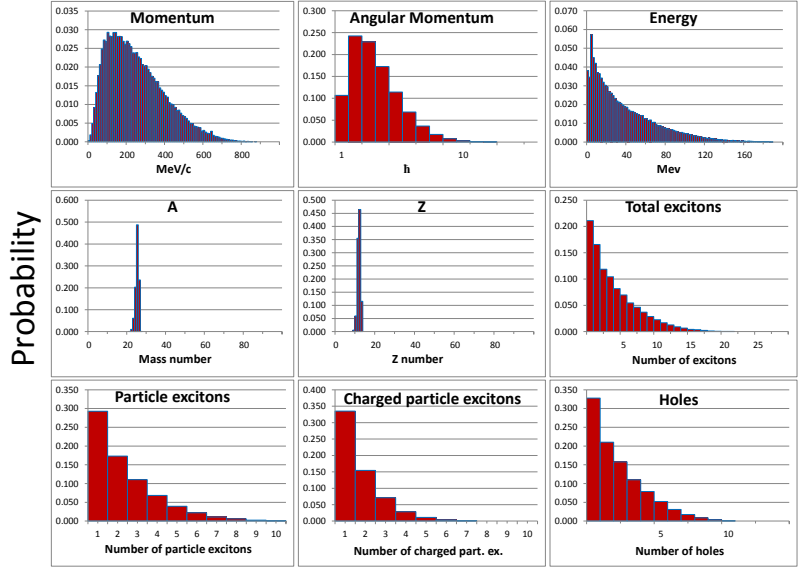}
\caption[]{Momentum, angular momentum, energy, mass- and charge-numbers, number of excitons, particle excitons, charged particle excitons, and holes distributions of residual nuclei for the 200 MeV p + $^{27}$Al $\rightarrow$ ... reaction directly after the INC, before the preequilibrium stage.}
\label{fig:p200AlHist}
\end{figure}

\begin{figure}[hbp]
\centering
\includegraphics[trim = 0in 1.0in 0in 1.0in, width=5.0in]{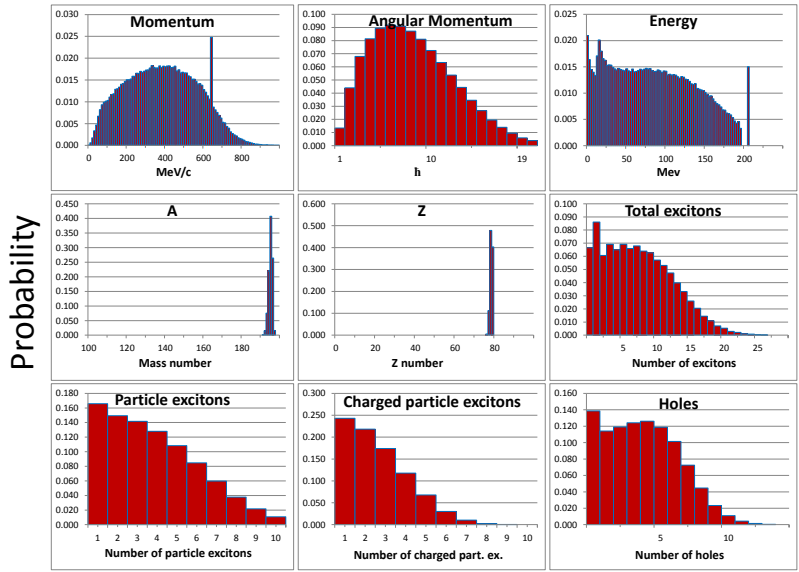}
\caption[]{Momentum, angular momentum, energy, mass- and charge-numbers, number of excitons, particle excitons, charged particle excitons, and holes distributions of residual nuclei for the 200 MeV p + $^{197}$Au $\rightarrow$ ... reaction directly after the INC, before the preequilibrium stage.}
\label{fig:p200AuHist}
\end{figure}

\begin{figure}
\centering
\includegraphics[trim = 0in 1.0in 0in 1.0in, width=5.0in]{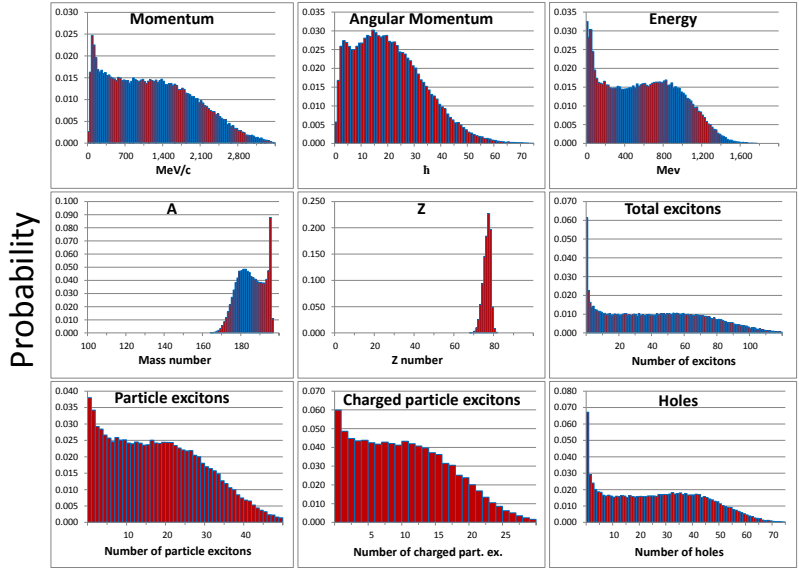}
\caption[]{Momentum, angular momentum, energy, mass- and charge-numbers, number of excitons, particle excitons, charged particle excitons, and holes distributions of residual nuclei for the 2500 MeV p + $^{197}$Au $\rightarrow$ ... reaction directly after the INC, before the preequilibrium stage.}
\label{fig:p2500AuHist}
\end{figure}

Figures~\ref{fig:p200AlHist}, \ref{fig:p200AuHist}, and \ref{fig:p2500AuHist} show distributions of several properties of the residual nuclei after the INC, and right before the preequilibrium stage, for several reactions. 

Observe how the number of charged particle excitons drops off sharply in Figure~\ref{fig:p200AlHist}. This demonstrates that we should expect the cross section to decrease dramatically as fragment size increases. We would also expect emission of LF from the MEM will be less of a factor in this reaction as it is in reactions with larger targets and/or higher incident energies.

Increasing the size of our target nucleus leads to a more gradual decline in our number of charged particle excitons in Figure~\ref{fig:p200AuHist}, and thus we would expect more emission from the MEM. Also notice the spikes in both the momentum and energy histograms. The momentum spike corresponds to the momentum of the incident proton. The energy spike, which occurs at about 207 MeV, corresponds to the reaction in which the proton and its full 200 MeV of energy is absorbed within the gold nucleus. This provides 200 MeV from the incident proton plus approximately 7 MeV from the binding energy, for a total of 207 MeV. This is not a violation of energy conservation because we change reference frames--from the incident proton with 200 MeV in the laboratory system to the nucleus center-of-mass system which receives an extra 7 MeV from the binding energy of the proton. We did not see this spike in the aluminum target because the Al nucleus is too small and the incident proton (or a created scatter particle) escapes the nucleus.

In Figure~\ref{fig:p2500AuHist} the peaks disappear again, because the energy has significantly increased and either the incident proton or one or several of the created scatter particles escapes the gold nucleus. Notice that the number-of-charged-particle-excitons probability does not begin to drop until after about 15. This means our MEM could emit a large fragment with high-energy from this high-energy reaction.

More physics information can be extracted from Figures~\ref{fig:p200AlHist}--\ref{fig:p2500AuHist}. For example, the momentum influences the angular distribution of emitted fragments: the greater the momentum the more forward-peaked the emitted fragments will be. In addition, most multifragmentation models require an energy of at least $\geq 4 $~MeV per nucleon. Inspection reveals that we would expect multifragmentation to pertain to the reaction 2500~MeV~p~+ $^{197}$Au~$\rightarrow$~... only. Furthermore, the angular momentum effects the probability of fission, with greater angular momentum leading to more fissions. Thus we would expect more fissions in the 2500~MeV~p~+ $^{197}$Au~$\rightarrow$~... reaction than at the lower energy of 200 MeV presented in Figure~\ref{fig:p200AuHist}. Lastly, distributions of A and Z reveal the number of collisions that occurred in the target-nucleus, with larger distributions resulting from more collisions.

\section{Fermi Break-up}
\subsection{Background}
We investigated the impact of expanding the threshold of the Fermi break-up model. In previous versions of CEM03.03, residual nuclei with $A \leq 13$ were sent to Fermi break-up. We raised this threshold to $A \leq 16$ and $A \leq 20$ and plotted comparisons for several reactions.

\begin{figure}[here]
\centering
\includegraphics[trim = 1in 1.5in 1.0in 1.0in, width=5.5in]{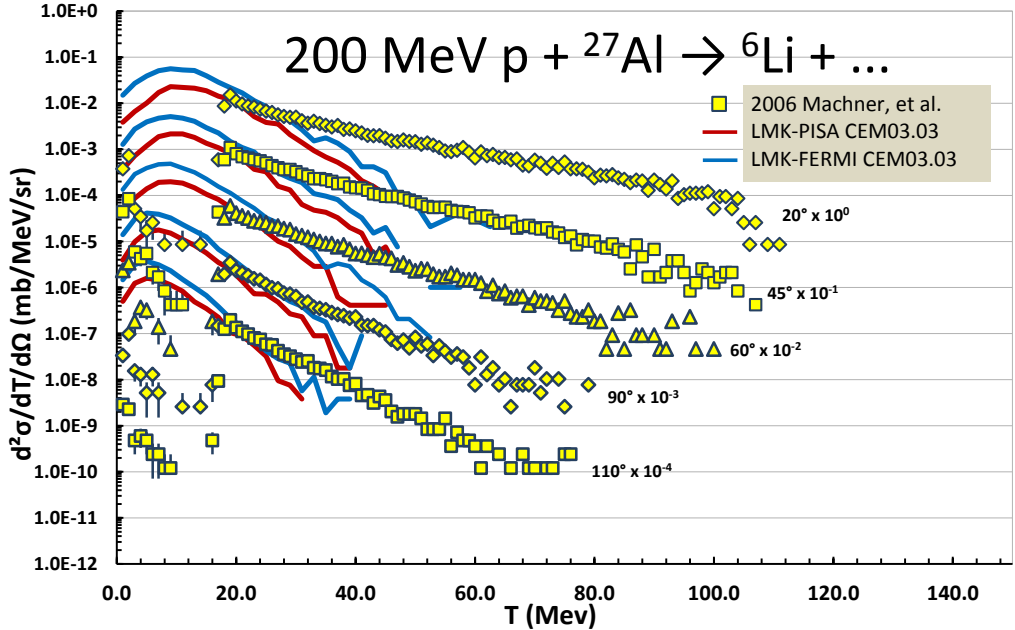}
\caption[]{Comparisons of experimental data by Machner et al. \cite{Machner} (open points) with results from the unmodified CEM03.03 (solid red lines) and the $A \leq 20$ Fermi-modified CEM03.03 (solid blue lines).}
\label{fig:FermiAlComp}
\end{figure}

Figure~\ref{fig:FermiAlComp} compares experimental data from Machner et al. \cite{Machner} with unmodified CEM03.03 and $A \leq 20$ Fermi-modified CEM03.03. As can be seen from this figure increasing the Fermi break-up cut-off does increase our cross sections of $^6$Li production by a relatively constant factor. However, this is not helpful in achieving better light-fragment cross sections at higher energies.

\begin{figure}
\centering
\includegraphics[trim = 0.5in 2.5in 1.0in 0.5in, width=6.5in]{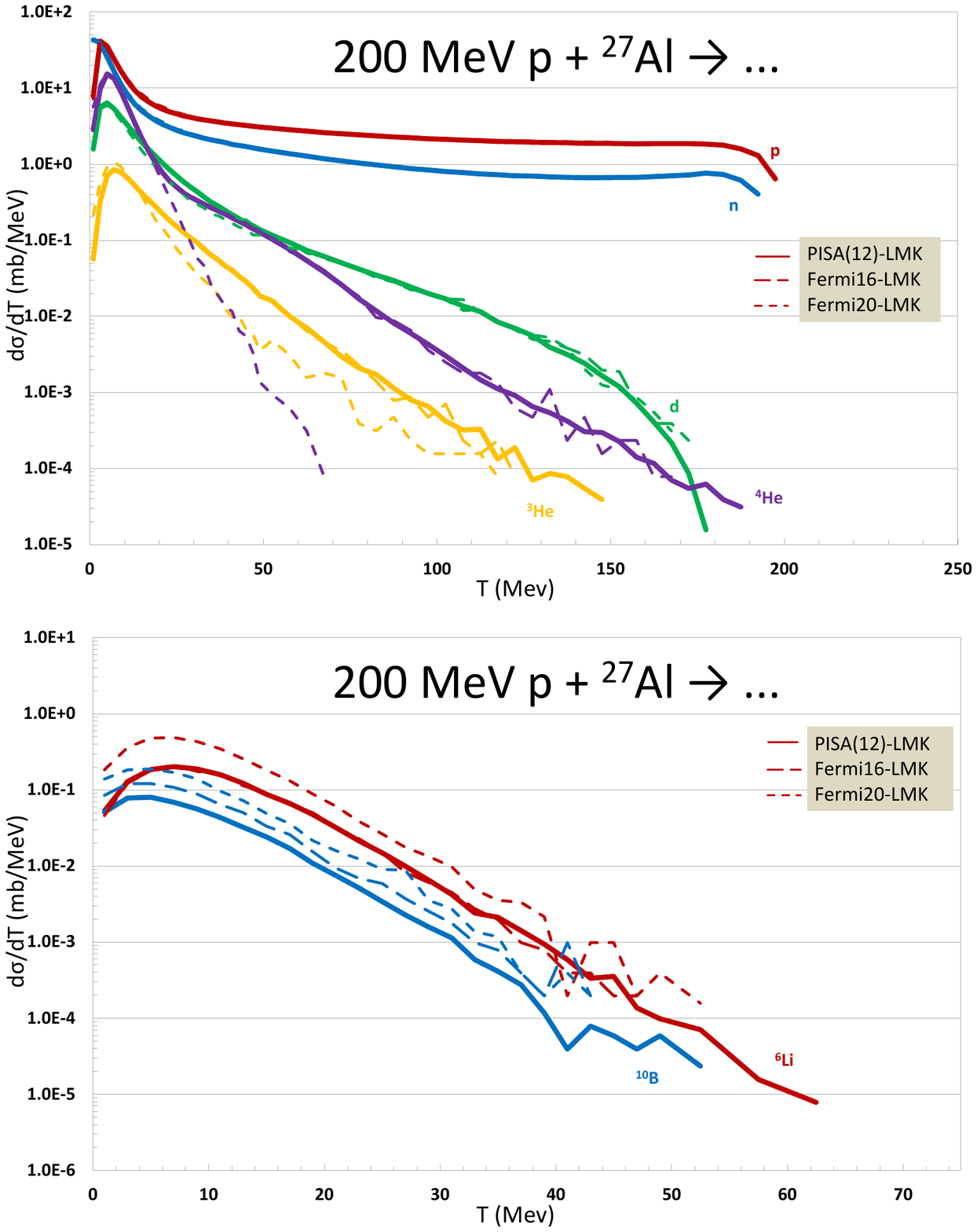}
\caption[]{Total angle-integrated cross sections for CEM03.03 (PISA(12)-LMK), $A \leq 16$ Fermi-modified CEM03.03 (Fermi16-LMK), and $A \leq 20$ Fermi-modified CEM03.03 (Fermi20-LMK).}
\label{fig:FermiAl}
\end{figure}

Figure~\ref{fig:FermiAl} shows that the cross sections of light fragments (lithium and beryllium) do, in fact, increase as the Fermi break-up cut-off is raised. However, with the cut-off at $A \leq 20$ we see significant deterioration in the $^4$He cross section at intermediate and high energies. This is unacceptable. A cut-off of $A \leq 16$ is acceptable, but if we wish to have a high cut off we would need to alter our Fermi break-up model to make its disintegration process more physical.

Attempts to increase the Fermi break-up cut-off above $A \leq 20$ resulted in fatal errors in the Fermi break-up code. This should be investigated and fixed. We should also return to adjusting the cut-off for Fermi break-up now that we have emission of light fragments in our MEM, to see how these adjustments affect cross sections at intermediate energies.

Further work with the Fermi break-up model is needed and planned for the future.

\subsection{Comparisons with Hagiwara et al. Experimental Data}
With the capability to output cross section by Z number we compared our unmodified CEM03.03 results to data recently published by Hagiwara et al. \cite{Hagiwara}.
\begin{figure}[htp]
\centering
\includegraphics[width=6.5in]{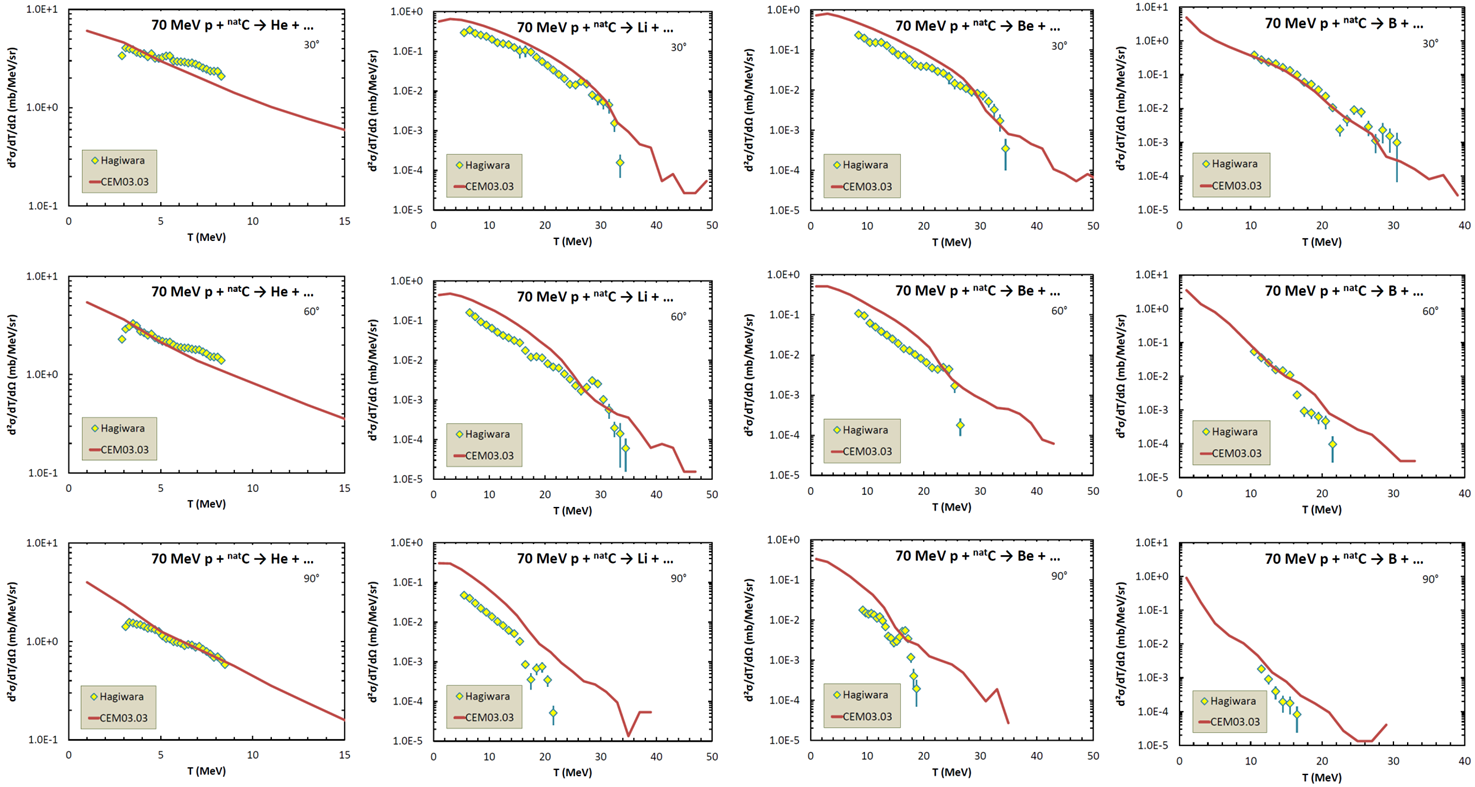}
\caption[]{Comparison of CEM03.03 (solid red lines) and experimental data by Hagiwara et al. \cite{Hagiwara} (open points) for a natural carbon target. Our calculations were performed for $^{12}$C.}
\label{Hagiwara}
\end{figure}

Figure~\ref{Hagiwara} demonstrates that CEM agrees reasonably well with Hagiwara's data for a natural carbon target, even for the light fragments Li, Be, and B, because for carbon targets precompound emission of light fragments is achieved through the Fermi break-up model (as $A$ is always $\leq 13$ for p + $^{12}$C). Hagiwara also has data for heavier silicon and aluminum targets, and in the future we plan to perform comparisons with this data and our new modified-MEM CEM03.03.

\end{document}